\documentclass[aps,prb,twocolumn,subeqn,groupedaddress,showpacs]{revtex4-1}
\usepackage{dcolumn}
\usepackage{epsfig}
\usepackage{amsmath}
\usepackage{graphicx}
\usepackage{subfigure}
\usepackage{color}
\begin{document}
\title{Unique Truncated Cluster Expansions for Materials Design \\ via  Subspace Projection and Fractional Factorial Design}
\author{Teck~L. Tan$^{1,2}$}\email{tantl@ihpc.a-star.edu.sg}
\author{D.~D. Johnson$^{3,4,1}$}\email{ddj@ameslab.gov,ddj@iastate.edu}
\affiliation{$^1$Department of Materials Science \& Engineering, University of Illinois, Urbana, IL 61801}
\affiliation{$^2$Institute of High Performance Computing, Agency for Science, Technology and Research, Singapore 138632, Singapore}
\affiliation{$^3$Ames Laboratory, U.S. Department of Energy, Iowa State University, Ames, IA 50011}
\affiliation{$^4$Department of Materials Science and Engineering, Iowa State University, Ames, IA 50011}

\date{\today}

\begin{abstract}
For alloy thermodynamics, we obtain unique, physical effective cluster interactions (ECI) from truncated cluster expansions (CE) via subspace-projection from a complete configurational Hilbert space; structures form a (sub)space spanned by a locally complete set of  cluster functions. 
Subspace-projection is extended using Fractional Factorial Design with subspace ``augmentation'' to remove systematically the ECI linear dependencies due to excluded cluster functions  --  controlling convergence and bias error, with a dramatic reduction in the number of structural energies needed.
No statistical fitting is required.
We illustrate the formalism for a simple Hamiltonian and Ag-Au alloys using density-functional theory.
\end{abstract}
\pacs{64.60.De, 64.60.Cn, 75.10.Hk, 02.30.Mv}
\maketitle


\section{Introduction}
A cluster expansion\cite{ClusterExpansion84} (CE) has proved to be an invaluable multi-scaling technique for generating cluster-based Hamiltonians that allow large numbers of configurational energies to be calculated efficiently via a small set of density functional theory \cite{Hohenberg1964, KohnSham1965} (DFT) calculations.
Hence, the CE Hamiltonian is well suited for modeling alloy thermodynamics and phase diagrams \cite{JPhaseEq23pg384, Garbulsky1994, PRB75p104203, AgAu2008, PRB67p064104, PRB36p4163} and perform
groundstate searches \cite {OrderingTend1995, SVBarabash06} over a large number of configurations on a fixed lattice.
Although the CE is an exact basis set expansion in terms of cluster correlation functions, \cite{ClusterExpansion84} whose coefficients (\textit{a priori} unknown) are effective cluster interactions (ECI), it is infeasible to determine all $\text{P}^\text{N}$ ECI for a P-component alloy on a N-site lattice (where N is large) as it would require the computation of all $\text{P}^\text{N}$ structural energies via expensive DFT calculations, defeating the purpose of CE as a multi-scaling tool.
For example, there are over 4 billion possible configurations for a a modest N=32 atom cubic cell of a (FCC) binary alloy (P=2).

{\par}
Instead, a truncated CE (trCE) is constructed from a training set of M$\ll$$\text{P}^\text{N}$  energies via structural inversion.\cite{ConnollyWilliams, JPhaseEq23pg384}
The truncation is, however, not unique -- but there is only one true set.
To minimize the mean-squared error associated with trCE, one has to balance the variance (data's numerical noise) and bias (inaccurate model for the estimator). 
Conventionally, ECI are treated as fitting parameters to obtain a trCE ``best fit" to known DFT structural energies (assumed random numerical noise).
To prevent over or under fitting, a `predictive' measure (e.g., leave-out-one cross-validation CV$_1$ error\cite{JPhaseEq23pg384}) is used to select a trCE, with emphasis on balancing errors from truncation and variance (data noise).
However, well-converged DFT energies should be virtually free of random noise.\cite{ECockayne_2010}
Also, for large learning sets, model selection via minimizing CV$_1$ could result in overfitting;\cite{Shao1993, Baumann2003395} trCE with CV$_1$ below the data's noise level should not be selected.\cite{PhysRevB.81.094116} Recent efforts to improve the predictive capability of trCE analyze only errors arising from variance.\cite{PhysRevB.80.165122, PhysRevB.82.184107}
Little has been done to address how bias impacts the predictive capability of trCE.

{\par} Here, with DFT structural energies assumed noiseless, we show that the only sources of error are the ECI of cluster functions excluded from the trCE set (the bias) and that the choice of structures in the set dictates the way errors are distributed.
Thus, the cluster functions included in the trCE can be linearly dependent on excluded cluster functions, affecting convergence and error.
We show that bias is reduced when physically important clusters are included in trCE.

{\par}The CE of a binary alloy (with complete cluster basis functions) is the Walsh-Hadamard transformation. \cite{OrthoArrays, Horadam2007} Combining this with concepts from fractional factorial design\cite{OrthoArrays, FactorialDesigns, Hierarchy2, NISTweb} (FFD) we show that linear dependencies between ECI can be deduced geometrically, if the M structures used in structural inversion are from a locally complete Hilbert subspace.
By prescribing a large supercell as the complete configuration Hilbert space, we can detail each subspace and identify linear dependencies between ECI.
In this \emph{subspace-projection}, structures form a (sub)space spanned by a \emph{locally complete set of cluster functions},\cite{SubspaceProjection} uniquely determined using a physical hierarchy.\cite{PRL92p55702}
Errors in trCE are eliminated when the subspace of known structures is large enough such that all physically significant ECI are included. 
What remains is the size of the subspace required and how one  resolves critical ECI linear dependencies.
To answer these, we use
\begin{enumerate}
\item{FFD concepts to construct complete (sub)spaces and identify linear dependencies between the excluded cluster functions and the truncated set.}
\item{Cluster hierarchy\cite{PRL92p55702}, established by the moment theorem,\cite{Cyrot1968, Cyrot1970, Cyrot1971} to ensure the choice of key physical ECI, yielding a locally complete CE that is unique.} 
\end{enumerate}
\noindent In addition, we elucidate the physical/mathematical meaning of ECI, which have at times been overlooked by treating the ECI only as fitting parameters, e.g., genetic algorithms searches,\cite{Hart2005} resulting in non-unique (often unphysical) clusters sets with similar CV scores.

{\par} We first review the CE formalism and its relationship to Hadamard matrices\cite{OrthoArrays} used in signal processing\cite{Horadam2007} and fractional factorial design\cite{OrthoArrays, FactorialDesigns, Hierarchy2, NISTweb} (or design of experiment), where the issues faced are similar to those in the CE.
Our subspace-projection CE formalism is illustrated first by a simple model Hamiltonian, and then by a detailed application to Ag-Au using DFT, where a subspace of modest size ($4$ times fewer energies vs. current methods) yields a trCE with good predictive capability without statistical fitting. The predictive capability of the trCE is validated with extra DFT structural energies and the ECI reflect the physical hierarchy used in subspace-projection. We note that statistical validation is applied in our methodology even though no statistical fitting is required.

\section{Cluster Expansion Overview} \label{Sec_CE}
The CE is a basis-set expansion of alloy properties in terms of cluster entities, giving a formal and exact representation \cite{ClusterExpansion84} when all clusters are included; in its most general form, the CE is applicable to any multi-component alloy on any fixed lattice.
Here  we use orthogonal cluster functions constructed from spin variables.

{\par}
Labeling the sites on an N-site lattice with integers $\{1, 2, ..., \text{N}\}$, the vector $\vec{\sigma}=\{\sigma_1, \sigma_2, ..., \sigma_{\text{N}}\}$ is used to describe a given structure (or configuration), where $\sigma_i=1(-1)$ if site $i$ is occupied by an A(B) atom in an A-B alloy.
Expanded in terms of cluster functions, the energy of an alloy structure is expressed as
\begin{align} \label{Eq_GCE}
  \text{E}(\vec{\sigma})=\sum_\eta J_{\eta} {\Phi}_\eta(\vec{\sigma})~~,
\end{align}
where $\eta=\{i_1, i_2, ..., i_n\}$ is a set of integers that denotes the sites selected to form an $n$-site cluster ($n\leq \text{N}$), with $i_k\in \{1, 2, ..., \text{N}\}$.
The summation is over all $2^\text{N}$ clusters possible within the N-site lattice, including $\eta_{0}=\{\emptyset\}$, which gives a constant term, $J_{_0}$, independent of $\vec{\sigma}$.
The $J_\eta$ coefficients are called the effective cluster interactions (ECI).
The cluster functions, $\Phi$, constructed from Chebyshev polynomials, \cite{ClusterExpansion84} are defined as
\begin{align}\label{Eq_Clus_Fn}
\Phi_\eta(\vec{\sigma})\equiv\sigma_{i_1}\sigma_{i_2}...\sigma_{i_n}, ~~\forall~~i_k\in \eta 
\end{align}
with $\Phi_{_{0}}$=$1$, and form an orthogonal basis set spanning the $2^{\text{N}}$ configuration space.
For example, $\Phi_{\{i\}}(\vec{\sigma})\equiv\sigma_i$ is the single-site function at site $i$ and  $\Phi_{\{i, j\}}(\vec{\sigma})\equiv\sigma_i\sigma_j$ is the pair function for sites $i$ and $j$, for a given $\vec{\sigma}$.
Note that, except for $\Phi_{_{0}}$, $\Phi_{\eta}(\vec{\sigma})$=$1$ or $-1$.

{\par}
Alternatively, \eqref{Eq_GCE} may be re-expressed in full to include all possible configurations and correlation functions.
\begin{align} \label{Eq_CE_Mat}
  \vec{\text{E}}{}=\left[\vec{\Phi}_1, \vec{\Phi}_2, ..., \vec{\Phi}_{2^\text{N}}\right]\vec{J}\equiv \mathbf{\Phi}\vec{J}~~.
\end{align}
$\vec{\text{E}}$ is a $2^\text{N}$-component vector with each component being the energy of one of the $2^\text{N}$ possible alloy structures, $\vec{\sigma}$; $\vec{\Phi}_\eta$ are $2^\text{N}$-component vectors, with each row being the correlation functions of structure $\vec{\sigma}$. Each cluster set $\eta$ is labelled by an integer from 1 to $2^\text{N}$.
The set of $\{\vec{\Phi}_\eta\}$ forms an orthogonal array and obeys the orthogonality condition,
\begin{align}\label{Eq_Clus_Ortho}
  \frac{1}{2^\text{N}} \vec{\Phi}_\eta \cdot \vec{\Phi}_{\eta'} \equiv \frac{1}{2^\text{N}}\mbox{Tr}^{(\text{N})}\Phi_\eta(\vec{\sigma}) \Phi_{\eta'}(\vec{\sigma})=\delta_{\eta\eta'}~~,
\end{align}
where the trace, $\mbox{Tr}^{(\text{N})}\equiv \sum_{\sigma_1} ...\sum_{\sigma_\text{N}}$, is over all $2^\text{N}$ configurations; this is essentially a dot-product between two correlation function vectors.

{\par}
If every  E$(\vec{\sigma})$ in $\vec{\text{E}}$ is evaluated (e.g., via DFT calculations),
the ECI are simply obtained from Eq. \eqref{Eq_CE_Mat} via a matrix inversion
\begin{align} \label{Eq_Inv_CE_Mat}
   \vec{J}=\mathbf{\Phi}^{-1}\vec{\text{E}}{}~~.
\end{align}
However, first-principle calculations are computationally costly, making it impossible to evaluate all E$^{\text{DFT}}(\vec{\sigma})$ for even a modest value of N (N=32 gives $\sim$ 4 billion configurations); thus, in practice, only a small fraction (typically between 30 to 100) are evaluated and used to construct a CE for an alloy system, and through structural inversion,\cite{ConnollyWilliams} only a subset of ECI can then be determined.
Therefore, two critical choices have to be made -- (1) the subset of E$^{\text{DFT}}$ for structural inversion and (2) the subset of ECI to be determined, which should not be left to guesswork.

\subsection{Error Analysis of Structural Inversion}\label{Sec_SI}
In the standard model for least-squares fitting, the observed values, $\vec{ \mathcal{E}}$, is related to the values of the true model, $\vec{\text{E}}$, by
\begin{align} \label{Eq_Noise}
  \vec{ \mathcal{E}}=\vec{\text{E}} + \vec{\epsilon}~~,
\end{align}
where $\vec{\epsilon}$ is a randomly distributed error with zero mean and variance $s^2$.
This implies that
\begin{align}
 \left< \vec{ \mathcal{E}}\right> = \vec{\text{E}}+\left<\vec{\epsilon}\right>=\vec{\text{E}}~~,
\end{align}
where $<...>$ denotes expectation values averaged over all possible observations having the same atomic configuration, $\vec{\sigma}$.  For us, the random noise in DFT data may arise from various computational setting (e.g., different energy-cutoff, k-points, convergent criteria). $< \mathcal{E}(\vec{\sigma})>$ is the expected energy of configuration $\vec{\sigma}$ averaged over all such computational settings. 

{\par}For a given $\vec{\sigma}$, the mean squared error (MSE) of its estimator, $ \hat{\text{E}}(\vec{\sigma})$, may be decomposed into a variance and bias,\cite{DataMiningBook} see Appendix \ref{Appendix}.
\begin{align}
 \text{MSE} &=  \left< \left(  \hat{\text{E}}(\vec{\sigma})-\text{E}(\vec{\sigma}) \right )^2  \right> \nonumber \\
 &=  \left<\left( \hat{\text{E}}(\vec{\sigma})- \left<\hat{\text{E}}(\vec{\sigma})\right>\right)^2\right>+ \left<\left(\left<\hat{\text{E}}(\vec{\sigma})\right>-\text{E}(\vec{\sigma}) \right)^2  \right> \nonumber \\
 &=\text{Var} + \text{Bias}~~.
\end{align}
$\text{E}(\vec{\sigma})$ is the true value and $< \hat{\text{E}}(\vec{\sigma})>$ is the estimator constructed from the expected observations. 

{\par} To minimize the MSE, one has to balance the variance (from data noise) and bias (inaccurate model for the estimator). For a fixed learning set, a trCE that includes too few clusters gives a large bias, although the variance maybe small (under-fitting), while too many clusters leads to over-fitting (large variance). Both under and over fitting lead to a large MSE and thus to a large prediction error.
To balance the variance and bias, most CE practitioners use CV$_1$, with issues \cite{PhysRevB.81.094116} discussed in the introduction. While the variance term is reduced with well-converged DFT energies and/or using more DFT energies (data points) in the fit, we show that the bias term is reduced when physically important clusters are included in the trCE.

{\par}We now re-write Eq. \eqref{Eq_CE_Mat} by dividing $\vec{J}$ into two subvectors $\vec{J}_1$ and $\vec{J}_2$ of length M and $2^\text{N}$-M, respectively, with $\vec{J}_1$ to be determined via structural inversion (SI), leaving out $\vec{J}_2$.
\begin{align} \label{Eq_CE_Mat_break}
  \left[ \begin{array}{c}
  \vec{\text{E}}_1\\
   \vec{\text{E}}_2\\
   \end{array} \right]
  =\left[
  \begin{array}{ccc}
  \boldsymbol{\phi}_{11} & \boldsymbol{\phi}_{12}\\
  \boldsymbol{\phi}_{21} & \boldsymbol{\phi}_{22}\\
  \end{array}
  \right] \left[
  \begin{array}{c}
    \vec{J}_{1}\\
    \vec{J}_{2}\\
  \end{array}
  \right]~~,
\end{align}
where $\boldsymbol{\phi}_{11}$ is a L-by-M matrix with $2^\text{N}\geq \text{L}\geq \text{M}$.
The variance term is then given by
\begin{align}
\text{Var}&={\phi}_{R1}^{\vec{\sigma}}\left(\boldsymbol{\phi}^\text{T}_{11}\boldsymbol{\phi}_{11}\right)^{-1} {\phi}_{R1}^{\vec{\sigma}~\text{T}} s^2 = \Lambda s^2
\end{align}
where ${\phi}_{R1}^{\vec{\sigma}}$ is a row vector of cluster functions in $\boldsymbol{\phi}_{R1}$ (where R=1 or 2) corresponding to configuration ${\vec{\sigma}}$ and $(\boldsymbol{\phi}^\text{T}_{11}\boldsymbol{\phi}_{11})^{-1}$ is the covariance matrix. Although the variance of the data noise is fixed at $s^2$ and beyond one's control, the variance term may be reduced by including specific configurations that will reduce $<\Lambda>_{\vec{\sigma}}$ (where $<...>_{\vec{\sigma}}$ is an average over a large set of configurations).\cite{PhysRevB.80.165122, PhysRevB.82.184107}

{\par}As for the bias term,
\begin{align}
\text{Bias}&=\left(\left<\hat{\text{E}}(\vec{\sigma})\right>-\text{E}(\vec{\sigma}) \right)^2 \nonumber \\
&=\left(  {\phi}_{R1}^{\vec{\sigma}}  \vec{\hat{J}}_1 - {\phi}_{R1}^{\vec{\sigma}}  \vec{{J}}_1 -  {\phi}_{R2}^{\vec{\sigma}} \vec{{J}}_2    \right)^2~~.
\end{align}
$\vec{\hat{J}}_1$ is the estimator of $\vec{J}_1$ and is obtained via SI using a least-squares method
\begin{align}\label{Eq_SI}
  \vec{\hat{J}}_1= \left(\boldsymbol{\phi}^\text{T}_{11}\boldsymbol{\phi}_{11}\right)^{-1} \boldsymbol{\phi}^\text{T}_{11}\vec{\text{E}}_1~~,
\end{align}
provided that $\boldsymbol{\phi}^\text{T}_{11}\boldsymbol{\phi}_{11}$ is invertible.
Detailed derivations for the variance and bias terms are in  Appendix \ref{Appendix}.

{\par}The choice of $\vec{\text{E}}_1$ already precludes certain combinations of $\vec{J}_1$ that would render $\boldsymbol{\phi}^\text{T}_{11}\boldsymbol{\phi}_{11}$ singular.
Notably, under the least-squares method, the estimator for $\vec J_2$ is always zero, i.e., $\vec{\hat{J}}_2=\vec 0$.
Unless $\vec{J}_2$ is truly zero, $\vec{\hat{J}}_1$ is a biased estimator; that is,
\begin{align} \label{Eq_BiasJ}
  \vec{\hat{J}}_1&= \vec{{J}}_1 + \left(\boldsymbol{\phi}^\text{T}_{11}\boldsymbol{\phi}_{11}\right)^{-1} \left( \boldsymbol{\phi}^\text{T}_{11}  \boldsymbol{\phi}_{12} \right) \vec{J}_2  
                            \equiv \vec{{J}}_1 + \delta \vec{J}_1~~,
\end{align}
derived by substituting $\vec{\text{E}}_1= \boldsymbol{\phi}_{11} \vec{J}_1+ \boldsymbol{\phi}_{12} \vec{J}_2$ from \eqref{Eq_CE_Mat_break} into \eqref{Eq_SI}.
The mean estimator of the known structural energies is thus
\begin{align} \label{Eq_LS_Error}
  \left< \vec{\hat{\text{E}}}_1\right>&= \boldsymbol{\phi}_{11}\vec{\hat{J}}_1 \nonumber \\
                             &=\vec{\text{E}}_1+\left[ \boldsymbol{\phi}_{11} \left(\boldsymbol{\phi}^\text{T}_{11}\boldsymbol{\phi}_{11}\right)^{-1} \left( \boldsymbol{\phi}^\text{T}_{11}  \boldsymbol{\phi}_{12} \right) - \boldsymbol{\phi}_{12} \right]\vec{J}_2 \nonumber \\
                             &\equiv  \vec{\text{E}}_1+\delta\vec{\text{E}}_1~~.
\end{align}
Likewise, for structural energies not used for SI, 
\begin{align}\label{Eq_Valid_Error}
   \left<\vec{\hat{\text{E}}}_2\right>&= \boldsymbol{\phi}_{21}\vec{\hat{J}}_1 \nonumber \\
                             &=\vec{\text{E}}_2+\left[ \boldsymbol{\phi}_{21} \left(\boldsymbol{\phi}^\text{T}_{11}\boldsymbol{\phi}_{11}\right)^{-1} \left( \boldsymbol{\phi}^\text{T}_{11}  \boldsymbol{\phi}_{12} \right) - \boldsymbol{\phi}_{22} \right]\vec{J}_2 \nonumber \\
                             &\equiv  \vec{\text{E}}_2+\delta\vec{\text{E}}_2~~.
\end{align}
Our goal is then to minimize the bias term over all structures, i.e., $<\text{Bias}>_{\vec{\sigma}}$, given by
\begin{align}
 <\text{Bias}>_{\vec{\sigma}}=\frac{|\delta\vec{\text{E}}_1|^2}{L} +\frac{|\delta\vec{\text{E}}_2|^2}{(2^N-L)}~~,
\end{align}
which will be achieved if $\vec{J}_2$=0, i.e., the true values of the excluded interactions are zero.
We stress that minimizing $|\delta\vec{\text{E}}_1|^2/L$ alone (i.e., least-squares fitting) will not minimize $<\text{Bias}>_{\vec{\sigma}}$  in general. In this case, a full rank invertible matrix $ \boldsymbol{\phi}_{11}$ would result in $|\delta\vec{\text{E}}_1|^2/\text{L}$=0.
However, unless $\vec{J}_2$=0, errors in $\vec {\text{E}}_2$ still remain
\begin{align}
\delta\vec{\text{E}}_2=\left[ \boldsymbol{\phi}_{21} \boldsymbol{\phi}_{11}^{-1}  \boldsymbol{\phi}_{12}  - \boldsymbol{\phi}_{22} \right]\vec{J}_2~~.
\end{align}
Thus, structures from $\vec{\text{E}}_2$ are needed for validation.


{\par} 
We thus see that the only source of error for the bias term comes from $\vec{J}_2$. Accepting that we have well-converged DFT energies, such that $\vec{\mathcal{E}}$ in \eqref{Eq_Noise} is precise and noiseless, one only needs to minimize $<\text{Bias}>_{\vec{\sigma}}$ to obtain a reliable trCE. We showcase an approach based on
fractional factorial design of experiments \cite{OrthoArrays, FactorialDesigns, Hierarchy2, NISTweb} to identify linearly dependent ECI and via a hierarchical approach, add physically important ECI to construct a unique trCE. In doing so, the number of physically important ECI in $\vec{J}_2$ decreases and one approaches the unique CE.  

We first show that errors are incurred when $\vec{\hat{J}}_1$ is evaluated with only a fraction of known ``experimental'' data  ($\vec{\text{E}}_1$). 
These concepts provide a specific method to select the structural energies for $\vec{\text{E}}_1$ such that $\boldsymbol{\phi}_{11}$ remains a Hadamard matrix and it shows clearly how $\vec{J}_2$ is the source of error for $\delta\vec{\text{E}}_1$,  $\delta\vec{\text{E}}_2$ and  $\delta\vec{J}_1$. 


\section{Relation to Hadamard Matrices}

When {$\{\vec{\Phi}_\eta\}$} in Eq. \eqref{Eq_CE_Mat} are arranged in a \emph{certain} lexicographical order,  $\mathbf{\Phi}$ becomes the Hadamard matrix, commonplace in factorial design \cite{FactorialDesigns} of experiments and signal processing \cite{Horadam2007}.
Several classes of Hadamard matrices exist, of which the Sylvester-type \cite{OrthoArrays} of size $2^\text{N}$-by-$2^\text{N}$ are of direct relevance to the CE.
Starting from a single lattice site labelled as 1,
\begin{align}
   \mathcal{H}_{\{1\}}=\left[
    \begin{array}{rr}
      1  & 1 \\
      1  & -1 \\
    \end{array}
    \right]
    =\left[\vec{\Phi}_0, \vec{\Phi}_{\{1\}}\right] ~~,
\end{align}
with the configuration space fully spanned by the 2-component vectors $\vec{\Phi}_0$ and $\vec{\Phi}_{\{1\}}$.
With two lattice sites,
\begin{align} \label {Eq_Ha2}
   \mathcal{H}_{\{1, 2\}}&=\mathcal{H}_{\{1\}}\otimes\mathcal{H}_{\{2\}}\equiv
    \left[
    \begin{array}{rr}
     \mathcal{H}_{\{2\}} & \mathcal{H}_{\{2\}}\\
     \mathcal{H}_{\{2\}} & -\mathcal{H}_{\{2\}}\\
    \end{array}
    \right] \nonumber \\
    &=\left[
    \begin{array}{rrrr}
      1  & 1 & 1 & 1 \\
      1  & -1 & 1 & -1 \\
      1  & 1  & -1 &  -1\\
      1 & -1  & -1 & 1 \\
    \end{array}
    \right] \nonumber \\
    &=\left[\vec{\Phi}_0, \vec{\Phi}_{\{1\}}, \vec{\Phi}_{\{2\}}, \vec{\Phi}_{\{1, 2\}}\right]~~.
\end{align}
The four possible configurations are given by $[\vec{\Phi}_{\{1\}}, \vec{\Phi}_{\{2\}}]$; e.g., the second row corresponds to a structure with atomic type $-1$ and $1$ on sites 1 and 2, respectively.
For a general N-site lattice the Hadamard matrix is 
\begin{align} \label{Eq_Full_Hadamard}
   \mathcal{H}_{\{1, ..., \text{N}\}}&=\mathcal{H}_{\{1\}}\otimes\mathcal{H}_{\{2\}} \otimes...\otimes \mathcal{H}_{\{\text{N}\}}~~,
\end{align}
which satisfies the property,
\begin{align}\label{Eq_Iden_Hada}
\mathcal{H}^\text{T}_{\{1, ..., \text{N}\}}\mathcal{H}_{\{1, ..., \text{N}\}}=2^\text{N}\mathbf{\text{I}}_{2^\text{N}}~~, 
\end{align}
where $\mathbf{\text{I}}_{2^\text{N}}$ is the $2^\text{N}$-by-$2^\text{N}$ identity matrix.
In addition, the columns and rows of $\mathcal{H}_{\{1, ..., \text{N}\}}$ are the Walsh functions, commonly used in spectral analysis of rectangular waveforms, \cite{Horadam2007} hence, in a complete $2^{\text{N}}$ vector space,
\begin{align} \label{Eq_phi_equal_H}
\mathbf{\Phi}= \mathcal{H}_{\{1, ..., \text{N}\}}~~.
\end {align}
Equations \eqref{Eq_CE_Mat} and \eqref{Eq_Inv_CE_Mat} are the Hadamard-Walsh transformation and its inverse, respectively, with the ECI being Walsh coefficients.

\section{Factorial Design and ECI of Isolated Cells} \label{Section_Fact_Design}
\subsection{ECI via Full Factorial Design}
The full factorial design space is spanned by the columns of the Hadamard matrix $\mathcal{H}_{\{1, ..., \text{N}\}}$.
Using N=2 for illustration and Eq. \eqref{Eq_Ha2}, the full factorial design is given by
\begin{align} \label {Eq_EHa2}
  \left[ E_{11}, E_{\bar{1} 1}, E_{1 \bar{1}},E_{\bar{1}\bar{1}} \right]^{\text{T}} = \mathcal{H}_{\{1, 2\}} \left[J_0, J_{1}, J_{2}, J_{1, 2} \right]^{\text{T}}~~,
\end{align}
where the subscripts of $\vec{\text{E}}$ denote the combination of $\vec{\sigma}$  (c.f. Eq. \eqref{Eq_Ha2}) while those of $\vec{J}$ label atomic sites, i.e., $i_n=1$ or 2. $J_{1}$ and $J_{2}$ are single-site interactions of site 1 and 2, respectively, while $J_{1,2}$ is the 2-body (pair) interaction between sites 1 and 2. We emphasize that such a formalism is identical to a CE of an isolated cell with no periodic boundary conditions (see  Fig. \ref{Fig_2AT_cell}).

In the nomenclature of factorial design,\cite{OrthoArrays, FactorialDesigns, Hierarchy2, NISTweb} $\vec{\text{E}}$ is called the full experimental \emph{data set} to be explained using N \emph{factors} (sites 1 and 2) with each factor having 2 possible  \emph{levels}, $1$ or $-1$ (analogous to the spin variable at each site). 
$\vec{\text{E}}$ consists of $2^\text{N}$ data points, with each represented by a unique \emph{combination} of N \emph{levels} (subscripts of  $\vec{\text{E}}$).
 $\vec{\text{E}}$ is fully explained by a model consisting of $2^\text{N}$ \emph{effects}, consisting of a constant, the N factors (single-site clusters) and all possible \emph {interactions} between the factors constructed by multiplying the relevant factors (pair and multibody clusters).

\subsection{Physical meaning of  $\vec{J}$}
The coefficients, $\vec{J}$, via the matrix inversion of $ \mathcal{H}_{\{1, 2\}} $, have specific physical meanings.
Specifically, 
\begin{align} 
       J_{0}&=&\frac{1}{4} (~E_{11}+E_{\bar{1} 1}~+~E_{1 \bar{1}}+E_{\bar{1}\bar{1}}~)  \label{Eq_J_value_1}  \\ 
      J_{1}&=& \frac{1}{4} (~E_{11}-E_{\bar{1} 1}~+~E_{1 \bar{1}}-E_{\bar{1}\bar{1}}~)    \\ 
      J_{2}&=& \frac{1}{4} (~E_{11}-E_{1 \bar{1}}~+~E_{ \bar{1} 1}-E_{\bar{1}\bar{1}}~)    \\
   J_{1,2}&=& \frac{1}{4} ([E_{11}-E_{\bar{1} 1}]-[E_{1 \bar{1}}-E_{\bar{1}\bar{1}}])~~. \label{Eq_J_value_4}
\end{align}
Here $J_{0}$ gives the average value of all 2$^2$ levels; $J_{1}$ gives the \emph{contrast} of effect 1 (single-site cluster at site 1) averaged over all possible levels of effect 2 (single-site cluster at site 2); that is, $E_{11}-E_{\bar{1} 1}$ and $ E_{1 \bar{1}}-E_{\bar{1}\bar{1}}$ measure the effect of the changing the levels in effect 1 with effect 2 fixed at levels $1$ and $-1$, respectively.
Likewise, $J_{2}$ gives the contrast of effect 2 averaged over all possible levels in effect 1. 
As for the 2-body interaction  $J_{1,2}$, the 1-body effects (given in square brackets) are contrasted with respect to each other.

{\par}
Thus, there is a clear \emph{physical meaning} and basis for the interactions, $J_{\eta}$.
As we see next, the numerical values of the ECI depends on how the cluster functions (the basis set) are chosen for truncation.

\subsection{ECI via Fractional Factorial Design}

{\par}
As noted in Section \ref{Sec_CE}, only a fraction of $2^\text{N}$ possible experimental data (this includes DFT structural energies) are obtained in practice, either because the experiments are costly or the total number required is prohibitively large.
In FFD, the \emph{sparsity of effects} (or Pareto's) principle \cite{Hierarchy2, EffectSparse} is assumed, i.e., all experiment data can be explained by a small number of effects.
In a 2-level FFD, $1/2^k$ ($k$ being an integer) of all possible experimental data are known.

{\par}
The FFD principle is useful for determining how the ECI in $\vec{J}$ are \emph{confounded};
i.e., how  interactions $\vec{J}_1$ and $\vec{J}_2$  in $\vec{J}$  (Eq.~\eqref{Eq_CE_Mat_break}) are correlated with one another.
[We use the accepted nomeclature ``confounded'', especially because it distinguishes basis-set truncation effects and actual physical correlations, e.g., atomic short-range order.]
Two ECI are confounded if it is impossible to ascertain their individual values from the known data set.
Specifically, if $\boldsymbol{\phi}_{11}$ is a Hadamard matrix, each ECI in $\vec{J}_2$ will be confounded with one and only one ECI in $\vec{J}_1$.

{\par}
Using \eqref{Eq_EHa2} as an example, suppose only the first two experiments, $E_{11}$ and $E_{\bar{1}1}$, are evaluated (half of the four possible experiments).
 This subset forms a combinatoric subspace where all possibilities of $\vec{\Phi}_{\{1\}}$  are included with $\vec{\Phi}_{\{2\}}$ held at a fixed value of 1.
As a result
\begin{align} \label {Eq_Frac2}
  \left[ E_{11}, E_{\bar{1}1} \right]^{\text{T}} = \left[
  \begin {array}{rrrr}
    1 & ~1 & 1 & ~1\\
    1 & -1&~1 & -1
  \end{array} \right]
  \left[J_0, J_{1}, J_{2}, J_{1, 2} \right]^{\text{T}}~~,
\end{align}
which is an under-determined set of linear equations; hence, it is impossible to solve for all ECI.
At best only two of the four can be determined.
Now, note, because columns 1 and 3 of the (effect) matrix are identical, $J_0$ and $J_{2}$ are confounded, and likewise for $J_{1}$ and $J_{1,2}$.
We now choose to evaluate two ECI, include them in $\vec{J}_1$ and evaluate via \eqref{Eq_SI}.
We avoid evaluating $J_0$ and $J_{2}$ or $J_{1}$ and $J_{1, 2}$, which would render $\boldsymbol{\phi}_{11}$ singular.

{\par} Now, suppose we choose to evaluate $J_{0}$ and $J_{1}$, comparing \eqref{Eq_CE_Mat_break} with \eqref{Eq_EHa2}, we have
\begin{align}
   \vec{\text{E}}_1=\left[  E_{11}, E_{\bar{1} 1} \right]^\text{T}~~&;~~\vec{\text{E}}_2=\left[  E_{1\bar{1}}, E_{\bar{1} \bar{1}} \right]^\text{T}~~, \\
   \vec{{J}}_1=\left[  J_{0}, J_{1} \right]^\text{T}~~&;~~\vec{{J}}_2=\left[ J_{2}, J_{1, 2} \right]^\text{T}~~,\\
   \boldsymbol{\phi}_{11}=\boldsymbol{\phi}_{12}=\boldsymbol{\phi}_{21}&=-\boldsymbol{\phi}_{22}= \mathcal{H}_{\{1\}}~~,
\end{align}
where the last relation results from the property of Hadamard matrices, see \eqref{Eq_Ha2}.
From Eq.~\eqref{Eq_Iden_Hada}, the error analysis for the truncated case in Eqs. \eqref{Eq_BiasJ}--\eqref{Eq_Valid_Error} are, respectively, simplified to
\begin{align}
   \vec{\hat{J}}_1&=\vec{J}_1+\vec{J}_2~~, \label{Eq_Hada2_confound} \\ 
   \delta\vec{\text{E}}_1&=0~~, \\
   \delta\vec{\text{E}}_2&=2\mathcal{H}_{\{1\}}\vec{J}_2~~. \label{Eq_Hada2_part3}
\end{align}
From \eqref{Eq_Hada2_confound}, the truncated set, $\vec{\hat{J}}_1$= [$J_{0}+J_2$,$J_{1}+J_{1, 2}]^{T}$, is clearly a bias estimate of $\vec{{J}}_1$.
While $\vec{\text{E}}_1$ is reproduced exactly, sources of error \eqref{Eq_Hada2_part3} for $\vec{\text{E}}_2$ lies \emph{solely} in $\vec{J}_2$.

{\par} To ensure that $\vec{\hat{J}}_1$ predicts structural energies well, the ECI in $\vec{J}_2$ should be zero (or negligible); however, these  values of $\vec{J}_2$ are not known \textit{a priori}.
A physics-based hierarchy is thus needed to rank the relative importance of the ECI.

\subsection{Hierarchical Order and Heredity Effect}
The crux of the issue is that one ECI from each of the two sets, $\{J_{0}, J_2\}$ and $\{J_{1}, J_{1, 2}\}$, has to be neglected, because the two ECI in each set is confounded.
In FFD, this choice is in general made using the \emph{hierarchical ordering} principle, \cite{Hierarchy1, Hierarchy2} i.e., higher-order interactions are smaller in magnitude and hence less important than lower-order ones.
With the \emph{a priori} assumption that $|J_0| >|J_{2}|$ and $|J_1| >|J_{1,2}|$,
one would choose to evaluate $J_0$ over $J_{2}$ and $J_1$ over $J_{1, 2}$.
In addition, the \emph{effect heredity} \cite{Hierarchy2, Heredity} principle states that if a higher-order effect is important, then at least one of its parent effect is important.
Thus, if we had instead evaluated $J_{1, 2}$, both $J_{1}$ and $J_{2}$ (parent effects of $J_{1, 2}$) must be evaluated.

{\par}
The concepts from FFD are necessarily applicable to CE.
Because one only evaluates the DFT energies of a fraction of all $2^\text{N}$ configurations, the ECI are confounded; and the confounding relations are affected by the choice of structures in $\vec{\text{E}}_1$.
When the structures form a complete configuration subspace, we show below that the confounding relation can be explained via geometry.
A physical hierarchy \cite{PRL92p55702} is used to select the physically most important ECI from a set of confounded ECI; clusters with less number of sites and smaller spatial extent are physically more important.
When the trCE basis is compact and locally complete the effect heredity principle is observed as well.
Notably, such principles also reflect the underlying physical origin of the ECI in the CE, where a clear hierarchy of clusters exists, \cite{PRL92p55702} as quantifiable from the moment theorem, \cite{Cyrot1968, Cyrot1970, Cyrot1971} which is  the fundamental basis for tight-binding (or Debye-H\"uckel) and the generalized perturbation methods.

\section{Factorial Design and ECI from Cells with Periodic Boundaries}

For the CE to represent correctly the thermodynamics of bulk alloys, the trCE has to be based on structures (or configurations) on an infinitely repeating lattice ($\text{N}\rightarrow\infty$). 
Hence, structural energies are calculated using periodic boundary conditions.
Typically in CE, the clusters and configurations are classified according to the underlying symmetry of the lattice.
The number of symmetry unique structures generated by an N-site lattice equals the number of symmetry--distinct clusters needed in the exact CE.
For structures, only the symmetry unique ones require evaluation via DFT, where methods exist to distinguish  symmetry unique ones. \cite{PRB75p104203, Hart2009}
When the clusters are classified according to symmetry (under the labels $n$ and $f$), the CE in Eq. \eqref{Eq_GCE} can be re-expressed as 
\begin{align} \label{Eq_Sym}
  \frac{\text{E}(\vec{\sigma})}{\text{N}}=\sum_{n, f} D_{nf}{J}_{nf} \left<{\Phi}_{nf}\right>_{\vec{\sigma}}~~,
\end{align}
where $n$ is the number of sites defining the cluster (e.g., $n$=2 for pairs) and $f$ enumerates symmetry--distinct clusters with the same $n$ but different spatial extent \cite{PRL92p55702} (e.g., for pairs, $f$=1 for nearest neighbor (NN) and $f$=2 for 2$^{nd}$ NN) and there are $D_{nf}$ degenerate clusters for each group (e.g., $D_{21}$=12 and  $D_{22}$=6 for the FCC Bravais lattice).

{\par}Clusters with the same label have the same ECI and the cluster function is averaged over all lattice sites, i.e.,
\begin{align} \label{Eq_Ave_Cor}
   \left<{\Phi}_{nf}\right>_{\vec{\sigma}}= \frac{1}{\text{N}} \sum_{i_1}^\text{N}\frac{1}{nD_{nf}}\sum_d^{D_{nf}}   {\Phi}_{\eta_{nfd}}(\vec{\sigma})~~,
\end{align}
where $\eta_{nfd}$ is the set of lattice sites $\{i_1, ..., i_n\}$ of a degenerate $n$-site cluster grouped under $n$,  $f$.
For a periodic structure, the site averaging is done in a  finite-sized unit cell. 
Hence, for a complete space and assuming S symmetry distinct clusters, the correlation matrix (see \eqref{Eq_CE_Mat} and \eqref{Eq_phi_equal_H}) is simplified into a $2^{\text{N}}$-by-S matrix,
\begin{align} \label{Eq_CE_Mat_Sym}
 \vec{\text{E}}{}=\left[\left<\vec{\Phi}_{11}\right>, ...,  \left<\vec{\Phi}_{1 f_\text{max}}\right>, ..., \left<\vec{\Phi}_{N f_\text{max}}\right>\right]\vec{J}'~~,
\end{align}
where each column vector is an average of columns in the  $2^{\text{N}}$-by- $2^{\text{N}}$ Hadamard matrix corresponding to the same cluster symmetry.
Truncating the cluster function basis set inherently confound ECI, as discussed above.
However, when truncating in a finite-sized Hilbert space that is periodically repeated, the ``confounding relations'' for the ECI can be deduced from geometry, as we  illustrate.

\begin{figure}
  \begin{center}
  \includegraphics[scale=0.5]{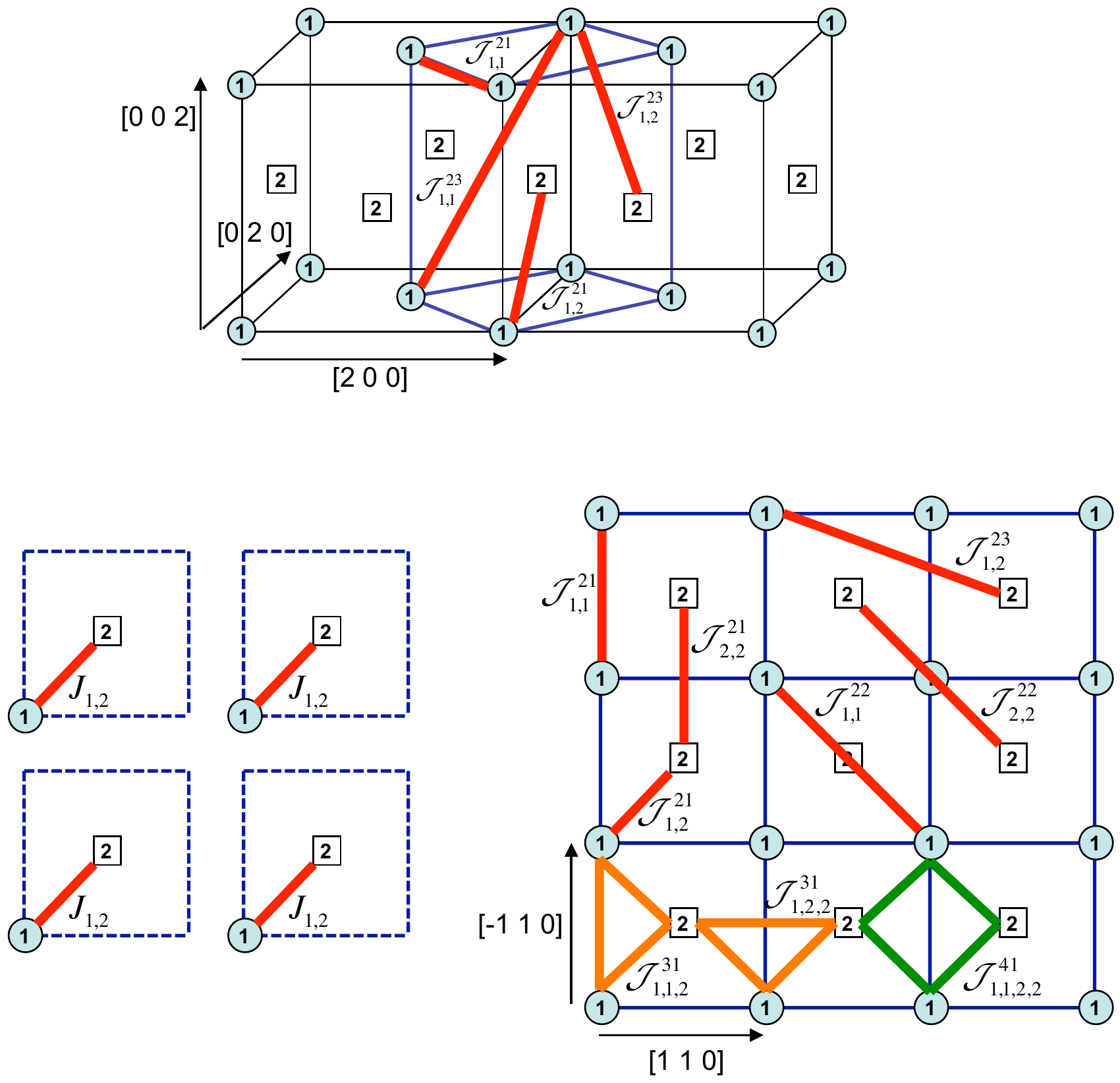}
   \caption{\label{Fig_2AT_cell} (color online) FCC lattice viewed in 3-D (top) and along [0 0 1]  (bottom right). The 2-site (4-site cubic) supercell is given by translation vectors $[1~1~0], [-1~{1}~0]$ and $[0~0~2]$  ($[2~0~0], [0~2~0]$ and $[0~0~2]$). Isolated 2-site cells are shown with dashed line (bottom left). The cell is periodically repeated to form an FCC lattice (bottom right). Some ECI (see text) are highlighted with bold lines for pairs (red), 3-body (orange) and 4-body (green). For clarity, selected pair ECI are highlighted on the top figure as well.  } 
\end{center}
\end{figure}

\subsection{Confounding Relations between ECI}\label{Sec_Confound}
We illustrate the confounding relations between the ECI by considering, for simplicity, a 2-site FCC supercell defined by translation vectors $[1~1~0], [-1~1~0], [0~0~2]$, see  Fig. \ref{Fig_2AT_cell}.
The single-site cluster function, $\Phi_{\{i\}}\equiv \sigma_i$, at site $i\in \{1, 2\}$ is $1$ ($-1$) if occupied by B (A).
A complete configuration space is formed if all 2$^2$ states in the supercell are considered.

We start by considering the 2-atom supercell to be isolated (infinitely separated from other supercells).
In this case, Eq. \eqref{Eq_EHa2} constitutes the full Hilbert space of the isolated cell, where the conceivable interactions include only a constant, two single-site and one pair term, see Fig. \ref{Fig_2AT_cell}.
Eq. \eqref{Eq_EHa2} can be re-written as
\begin{align} \label {Eq_IsolatedCell}
  \vec{\text{E}}  = J_0 \vec{\Phi}_{\{0\}}+J_{1} \vec{\Phi}_{\{1\}}+ J_{2} \vec{\Phi}_{\{2\}}+ J_{1, 2}  \vec{\Phi}_{\{1,2\}}~~,
\end{align}
where $\vec{\Phi}_{\eta}$ ($\eta={\{i_1,...,i_n\}}$) are defined in \eqref{Eq_Ha2}.
When the supercells are assembled to form the FCC lattice, many interaction terms are possible (Fig. \ref{Fig_2AT_cell}) and the energy of each 2-atom supercell is given by
\begin{align} \label {Eq_BC_Cell}
  \vec{\text{E}}  = & ~\mathcal{J}_0^{01} \vec{\Phi}_{\{0\}}+\mathcal{J}_{1}^{11} \vec{\Phi}_{\{1\}}+ \mathcal{J}_{2}^{11} \vec{\Phi}_{\{2\}}+ 8\mathcal{J}_{1, 2}^{21}  \vec{\Phi}_{\{1,2\}} \nonumber  \\
  &+ 2\mathcal{J}_{1,1}^{21}\vec{\Phi}_{\{0\}}+2\mathcal{J}_{2,2}^{21}\vec{\Phi}_{\{0\}}+3\mathcal{J}_{1,1}^{22}\vec{\Phi}_{\{0\}}+3\mathcal{J}_{2,2}^{22}\vec{\Phi}_{\{0\}} \nonumber \\
  &+ 4\mathcal{J}_{1,1}^{23}\vec{\Phi}_{\{0\}}+ 4\mathcal{J}_{2,2}^{23}\vec{\Phi}_{\{0\}}+16\mathcal{J}_{1,2}^{23} \vec{\Phi}_{\{1,2\}} + ... \nonumber\\
  &+ 8\mathcal{J}_{1,1,2}^{31}\vec{\Phi}_{\{2\}}+8\mathcal{J}_{1,2,2}^{31}\vec{\Phi}_{\{1\}} + ... \nonumber \\
  &+ 4\mathcal{J}_{1,1,2,2}^{41}\vec{\Phi}_{\{0\}}+ ... \nonumber \\
  &+ ...~~.
\end{align}
The interaction superscripts of $\mathcal{J}_{\eta}^{nf}$ are the symmetry indices used for the CE, see Eq. \eqref{Eq_Sym}, and the numerical prefactor gives the cluster degeneracy based on the symmetry at each of the two atomic site. 
For clarity, we have limited the expansion to only the nearest neighbor (NN) multibody ECI.

{\par} The confounding relations between $\mathcal{J}_{\eta}^{nf}$ are apparent when \eqref{Eq_BC_Cell} is written as
\begin{align} \label {Eq_BC_Cell_2}
  \vec{\text{E}}  = 
 &~~~[ \mathcal{J}_0^{01}+2\mathcal{J}_{1,1}^{21}+2\mathcal{J}_{2,2}^{21}+3\mathcal{J}_{1,1}^{22}+3\mathcal{J}_{2,2}^{22} \nonumber\\
 &+ 4\mathcal{J}_{1,1}^{23}+4\mathcal{J}_{2,2}^{23}+ ... +4\mathcal{J}_{1,1,2,2}^{41}+...] \vec{\Phi}_{\{0\}} \nonumber \\  
 &+[\mathbf{\mathcal{J}}_{1}^{11}+8\mathcal{J}_{1,2,2}^{31}+...] \vec{\Phi}_{\{1\}} \nonumber  \\
 &+[\mathbf{\mathcal{J}}_{2}^{11}+ 8\mathcal{J}_{1,1,2}^{31}+...] \vec{\Phi}_{\{2\}} \nonumber  \\
 &+[8\mathbf{\mathcal{J}}_{1, 2}^{21}+16\mathcal{J}_{1,2}^{23} ...]  \vec{\Phi}_{\{1,2\}}~~.
\end{align}
Based on a physical hierarchy, $\mathcal{J}_0^{01}$, $\mathcal{J}_{1}^{11}$, $\mathcal{J}_{2}^{11}$ and $\mathcal{J}_{1, 2}^{21}$ are ECI of the most compact and important clusters and they are not confounded. Each of them is however confounded with other ECI of larger spatial extent and larger $n$.
The confounding relations are conveniently revealed by annihilating repeated subscripts (which denotes atomic sites) of $\mathcal{J}_{i_1,...,i_n}^{nf}$ i.e.,
 \begin{align}\label{Eq_shorthand_B} 
  i_n, i_m&\rightarrow \{\emptyset\}, ~\text{if}~ i_n=i_m~~.
\end{align}
Interactions with the same irreducible subscripts will be confounded.
For example, using \eqref{Eq_shorthand_B} the subscripts of $\mathcal{J}_{1,2,2}^{31}$ leads to 1 and, hence, the interaction is confounded with $\mathcal{J}_{1}^{11}$. 
Note that $\mathcal{J}_0^{01}$, $\mathcal{J}_{1}^{11}$, $\mathcal{J}_{2}^{11}$ and $\mathcal{J}_{1, 2}^{21}$, the compact and physically most important ECI, have irreducible subscripts for the 2-atom supercell.

\subsection{Additional Symmetry Constraints}\label{Sec_Add_Const}
As pointed out earlier, for a cluster expansion on a Bravais lattice, e.g., FCC, symmetry degenerate clusters are grouped together as they have the same interaction values.
Hence, $\mathcal{J}_{\eta}^{nf}=J_{nf},~\forall~\eta$. 
For example, for point ECI $\mathcal{J}_{1}^{11}=\mathcal{J}_{2}^{11}=J_{11}$, and, for pair ECI, $\mathcal{J}_{1,1}^{21} = \mathcal{J}_{2,2}^{21} =\mathcal{J}_{1,2}^{21} =J_{21}$.
Enforcing the FCC symmetry on the 2-atom supercell generates only three unique structures (i.e., AA, AB and BB) and only three symmtery-unique ECI can be evaluated.
Eq. \eqref{Eq_BC_Cell_2} is then re-written as
\begin{align} \label{Eq_BC_Cell_3}
  \vec{\text{E}}  = &~[J_{0}+4J_{21}+6J_{22}+8J_{23}...+4J_{41}+...] \vec{\Phi}_{\{0\}} \nonumber \\  
 &+[J_{11}+8J_{31}+...] [\vec{\Phi}_{\{1\}}+\vec{\Phi}_{\{2\}}] \nonumber  \\
 &+[ 8J_{21}+16J_{23} ...]  \vec{\Phi}_{\{1,2\}}~~.
\end{align}
Thus, three distinct sets of confounded ECI exist, namely, those confounded with the
\begin{enumerate}
\item{constant $J_{0}$: $\{J_{0}, J_{21}, J_{22}, J_{23}, ..., J_{41}, ...\}$ }
\item{point $J_{11}$: $\{J_{11}, J_{31}, ...\}$ }
\item{1$^{st}$-NN pair $J_{21}$: $\{J_{21}, J_{23}, ...\}$ }
\end{enumerate}
The ECI are listed according to physical hierarchy in each group.\cite{PRL92p55702}

{\par}\emph{The Key Outcome} -- Selecting the most physically important cluster from each confounded set, $J_{0}$,  $J_{11}$ and  $J_{21}$ constitute the truncated (physical) $\vec{J}_1$,
which are the 3 independent ECI that can be determined within the present small Hilbert space. 
The neglected ECI bias the estimated $\vec {\hat{J}}_1$ according to \eqref{Eq_BiasJ}.
Using the orthogonality of the Hadamard matrix, the estimators of the ECI can be evaluated accordingly from \eqref{Eq_BC_Cell_3}
\begin{align} 
   \hat{J}_{0}+4\hat{J}_{21}&= \vec{\text{E}} \cdot \vec{\Phi}_{\{0\}} \nonumber \\
       &=\frac{1}{4} (~E_{11}+E_{\bar{1} 1}~+~E_{1 \bar{1}}+E_{\bar{1}\bar{1}}~)  \label{Eq_eval_J01} \\
   \hat{J}_{11}&= \vec{\text{E}} \cdot \vec{\Phi}_{\{1\}} = \vec{\text{E}} \cdot \vec{\Phi}_{\{2\}}  \label{Eq_eval_J11}\\
   8\hat{J}_{21}&= \vec{\text{E}} \cdot \vec{\Phi}_{\{1,2\}} \label{Eq_eval_J21}~~,
\end{align} 
with expressions similar to those in \eqref{Eq_J_value_1} to \eqref{Eq_J_value_4}.
From \eqref{Eq_eval_J11} it can be shown that $E_{\bar{1} 1}=E_{1 \bar{1}}$, implying the presence of only 3 unique structural energies, $E_{11}$, $E_{\bar{1}\bar{1}}$ and $E_{\bar{1} 1}$. 
Hence, the symmetry of the problem is properly reflected.
Notice, because the value of $\hat{J}_{21}$ is determined via \eqref{Eq_eval_J21}, $\hat{J}_{21}$ is no longer confounded with $\hat{J}_{0}$ in \eqref{Eq_eval_J01}.

{\par} Lastly, when the energy is normalized with respect to the number of atoms ($N=2$ in this case), the symmetry-reduced CE formalism given in \eqref{Eq_Sym} is recovered, i.e.,
\begin{align} \label {Eq_BC_Cell_4}
 \frac{\vec{\text{E}}}{2}  = &~\frac{J_{0}}{2}[\vec{\Phi}_{\{0\}}] +J_{11} [\frac{1}{2} \vec{\Phi}_{\{1\}}+\frac{1}{2}\vec{\Phi}_{\{2\}}]\nonumber \\
 &+ 12J_{21}[ \frac{2}{12}\vec{\Phi}_{\{0\}}+\frac{4}{12}\vec{\Phi}_{\{1,2\}}] +6J_{22}  [\frac{3}{6}\vec{\Phi}_{\{0\}}] \nonumber \\ 
 &+24J_{23} [\frac{4}{24} \vec{\Phi}_{\{0\}}+\frac{8}{24}\vec{\Phi}_{\{1,2\}}] + ...\nonumber \\
 &+24J_{31}[\frac{4}{24}\vec{\Phi}_{\{1\}}+\frac{4}{24}\vec{\Phi}_{\{2\}}] + ...\nonumber  \\
 &+ 8J_{41}[ \frac{2}{8} \vec{\Phi}_{\{0\}}] +... \nonumber \\
 =&~J_{01}\left<\vec{\Phi}_{01}\right>+J_{11}\left<\vec{\Phi}_{11}\right>\nonumber \\
 &+12J_{21}\left<\vec{\Phi}_{21}\right>+6J_{22}\left<\vec{\Phi}_{22}\right>+24J_{23}\left<\vec{\Phi}_{23}\right> +...\nonumber \\
 &+24J_{31}\left<\vec{\Phi}_{31}\right>+...+ 8J_{41}\left<\vec{\Phi}_{41}\right>+...~~,
\end{align}
where we have used Eq.~\eqref{Eq_Ave_Cor}. Notably, $\left<\vec{\Phi}_{01}\right>$ is a (constant) column vector of "1's".

\subsection{Physical Hierarchy of Clusters} \label{subsect-hierarchy}
As discussed earlier, $\vec{J}$ have physical meaning and are the coefficients of a Hadamard-Walsh transformation.
The importance of a cluster can be ranked according to the number of sites (order $n$) and spatial extent (range $f$).
The need of hierarchical arrangement is clear, without which one could equally likely choose to evaluate $J_{22}$, $J_{31}$ and $J_{23}$, see \eqref{Eq_BC_Cell_3}, and still obtain a solution because these ECI  are not confounded.
From the moment theorem, \cite{Cyrot1968, Cyrot1970, Cyrot1971} higher-order clusters are less important (smaller in magnitude), as verified in DFT. \cite{AgAu2008,GPMvsCE-PRB39p2242}

{\par} 
In addition, when an ECI of a higher-order cluster is included in $\hat{J}_1$, all ECIs belonging to its subclusters must also be included to give a locally complete CE set, \cite{PRL92p55702} as reflected in the heredity principle in FFD.
For clusters with the same $n$, those with larger spatial extent are less important, \cite{Cyrot1968, Cyrot1970, Cyrot1971,GPMvsCE-PRB39p2242} e.g., $|J_{21}| \gtrsim |J_{22}| \gtrsim |J_{23}| \gtrsim ...$~.
Together, these mathematical/physical criteria permit a hierarchy of ranges for $n$-body ECI, \cite{PRL92p55702} i.e., $r_{(n)}\geq r_{(n+1)}$; 2-body ECI are longer range than 3-body, which are longer range than 4-body, and so on.
Essentially, physical hierarchy \cite{PRL92p55702} requires that
\begin{enumerate}
\item{Higher-order (large $n$) clusters  are less important (but, if an $n$-body cluster is included in the trCE, its subclusters must be included).}
\item{For fixed $n$, clusters with larger spatial extent are less important.}
\end{enumerate}
These rules  maintain completeness within the local CE basis when mathematically implemented.

\subsection{Systematically Unconfounding Key ECI} \label{Sec_4AT_Cell}
As is clear from above, for a finite-size supercell we can group confounded ECI together utilizing concepts from FFD. 
The ECI in each confounded group are arranged according to a physical hierarchy and the most physically important ECI from each group is evaluated.
The structures in the chosen supercell thus constitutes a configuration subspace, which is necessarily spanned by the cluster functions of the most physically important ECI.
The key task is then to find a minimal subspace to achieve this, given that the number of ECI required for a general alloy is finite and follows the physical hierarchy.
To this end we offer the following resolution:
\begin{enumerate}
\item Define a large N-site supercell (with all possible $2^\text{N}$ configurations) as the ``complete'' Hilbert space.
\item  Select a reasonably-sized supercell in the Hilbert space as the initial subspace and $\vec{E}_1$ (see \eqref{Eq_SI}) includes DFT structural energies in this subspace.
\item From the physical hierarchy for clusters, the most important unconfounded ECI are evaluated via \eqref{Eq_SI}.
\item  Augment the subspace to unconfound key, longer-ranged ECI (especially pairs).  
That is, check for physically important clusters whose ECI remain confounded and then unconfound each targeted ECI by adding a structure systematically -- augmenting -- from the complete Hilbert space (not in the initial subspace) to $\vec{E}_1$. 
\end{enumerate}
The first step is a conceptual construct allowing us to define a large enough supercell as our complete space. When the ECI of all important cluster functions spanning this space is known (complete), the CE is able to predict accurately all structural energies of the alloy system.
A 2-atom supercell shown earlier is unlikely to unconfound  key ECI of a binary, so steps 2 and 3 have to be accomplished using a bigger supercell, as we now exemplify. 

\begin{table}[]
\begin{center}
\caption{\label{Tab_modelH} Model ${J}_{nf}$ and their degeneracy $D_{nf}$ for a FCC lattice. The estimate, $\hat{J}_{nf}$, via \eqref{Eq_SI} is given for structures belonging to the subspace of a 2-site cell and the complete space given by the 4-site cubic cell.
  }
\begin{tabular}{c|c|c|c|c|c}
\hline \hline
& & &${J}_{nf}$ &\multicolumn {2} {c} {~~~$\hat{J}_{nf}$ } \\
\hline
$n$ & $f$ & $D_{nf}$  &Model &2-site cell&4-site cell \\
\hline
0 & 1 & 1  & 1& 0.8 & 1 \\
1 & 1 & 1  & -1& -0.2 & -1 \\
2 & 1 & 12  & 1& 1 & 1 \\
3 & 1 & 24  & 0.1& 0 & 0.1 \\
4 & 1 & 8  & -0.1& 0 & -0.1 \\
\hline
\end{tabular}
\end{center}
\end{table}

\begin{figure}
  \begin{center}
  \includegraphics[scale=0.55]{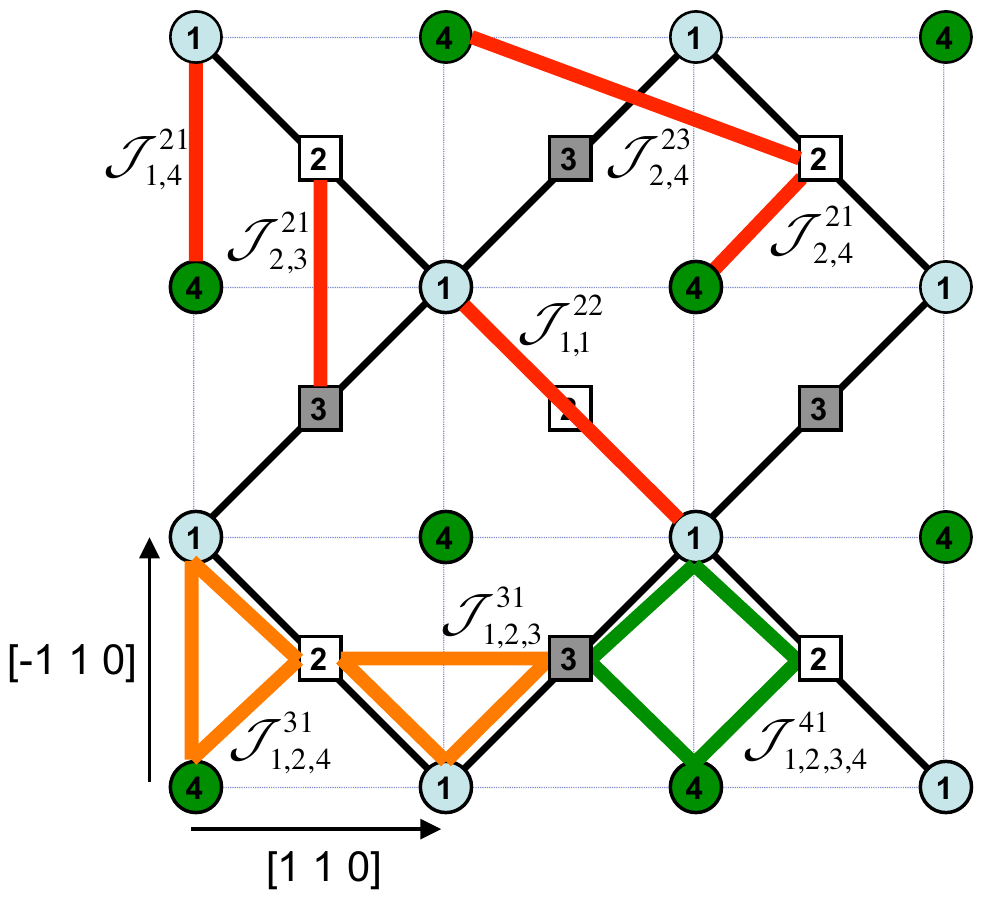}
   \caption{\label{Fig_4AT_cell} 
   (color online) 4-site supercells defining an FCC lattice (outlined in black) as viewed from the top, Fig. \ref{Fig_2AT_cell}. 
   ECI in  Fig. \ref{Fig_2AT_cell} are shown also. The subscripts of $\mathcal{J}_{\eta}^{31}$ and  $\mathcal{J}_{\eta'}^{41}$ are irreducible, \eqref{Eq_shorthand_B}, so $J_{31}$ and $J_{41}$ are no longer confounded with $J_{11}$ and $J_0$, respectively (see text). $J_{23}$ and $J_{21}$ are still confounded, as are $J_{22}$ and $J_{0}$.
    } 
\end{center}
\end{figure}

\subsubsection{An Illustrative Example}
{\par}For FCC binaries, we create a model CE Hamiltonian (values are in Table \ref{Tab_modelH}), such that all structural energies are defined by interactions within the nearest-neighbor (NN) range, i.e., only $J_{01}, J_{11}, J_{21}, J_{31}, J_{41}\neq 0$.
The model Hamiltonian is assumed to be unknown \textit{a priori} and we seek to estimate their values via \eqref{Eq_SI}.
We start by using the configuration subspace defined by the 2-site supercell (Fig. \ref{Fig_2AT_cell}) to calculate an estimator, $\vec{\hat{J}}_1=[\hat{J}_{01}, \hat{J}_{11}, \hat{J}_{21}]^\text{T}$, whose components are the most physically important ECI that are not confounded, see \eqref{Eq_BC_Cell_3}.
Because we did not span the complete space, $\vec{\hat{J}}_1$ is biased because some non-zero ECI will be confounded.
From \eqref{Eq_BC_Cell_3}, it is clear that $J_{01}$ will be confounded with $J_{41}$ while $J_{11}$ is confounded with $J_{31}$. 
This is indeed the case as shown in Table \ref{Tab_modelH}, e.g., 
\begin{align}\label{J0-confound}
\hat{J}_{01}=J_{01}+(D_{41}/4)J_{41} = 0.8 \ \ \ \ .
\end{align}
$J_{21}$ is not confounded with other non-zero ECI and is thus an unbiased estimate. 

{\par}We next consider the locally complete subspace generated by the cubic 4-site supercell, which contains all structures from the 2-site supercell and two new  (A$_3$B and AB$_3$) structures.
As illustrated in Fig. \ref{Fig_4AT_cell}, $J_{31}$ and  $J_{41}$ are no longer confounded with $J_{01}$ and $J_{11}$, respectively. 
This effectively unconfounds all non-zero ECI, so $\vec{\hat{J}}_1$ is an unbiased estimate of $\vec{{J}}_1$, as shown in Table \ref{Tab_modelH}.

{\par} Indeed, the confounding between ECI (due to the often arbitrary choices of cluster functions that are included) is the main cause of variation of ECI between different publications and different predictions, which can be now eliminated if we were lucky enough to choose a configurational subspace that is spanned by the cluster functions of all significant non-zero  $J_{nf}$.

{\par} For real alloy systems, $J_{2f}$ remains significant up to a longer range compared to multibody ECI ($n>2$), hence it is necessary to unconfound the ECI of longer range pairs. However, when the supercell is increased as shown in going from the 2-site to the 4-site cell, one discovers that only the NN multibody $J_{n1}$ ($n=3,~4$) are unconfounded, longer-ranged pairs such as $J_{22}$ and $J_{23}$ remains confounded, see Fig. \ref{Fig_4AT_cell} caption. 
The increase in supercell size unconfounds short-ranged multibodies at a faster rate than longer range pairs.
To unconfound longer-range pairs while keeping the number of required (DFT calculated) structural energies very small, structures from an augmented space (step 4 in the above resolution) are added systematically to the existing subspace. Importantly, one augmented structure is added at a time to unconfound a long range ECI.

{\par} Generally, unconfounded (unique) truncated ECI are achieve by a limited augmentation of the initial configuration subspace, as discussed in Sec~\ref{Sec_SSP} and illustrated for Ag-Au case study. We find that the truncated ECI from augmentation of the configuration subspace has comparable predictive capability as the one selected by CV$_1$ but with four times less structural energies.

\begin{figure}[]
  \begin{center}
  \subfigure[] {
  \includegraphics[scale=0.55]{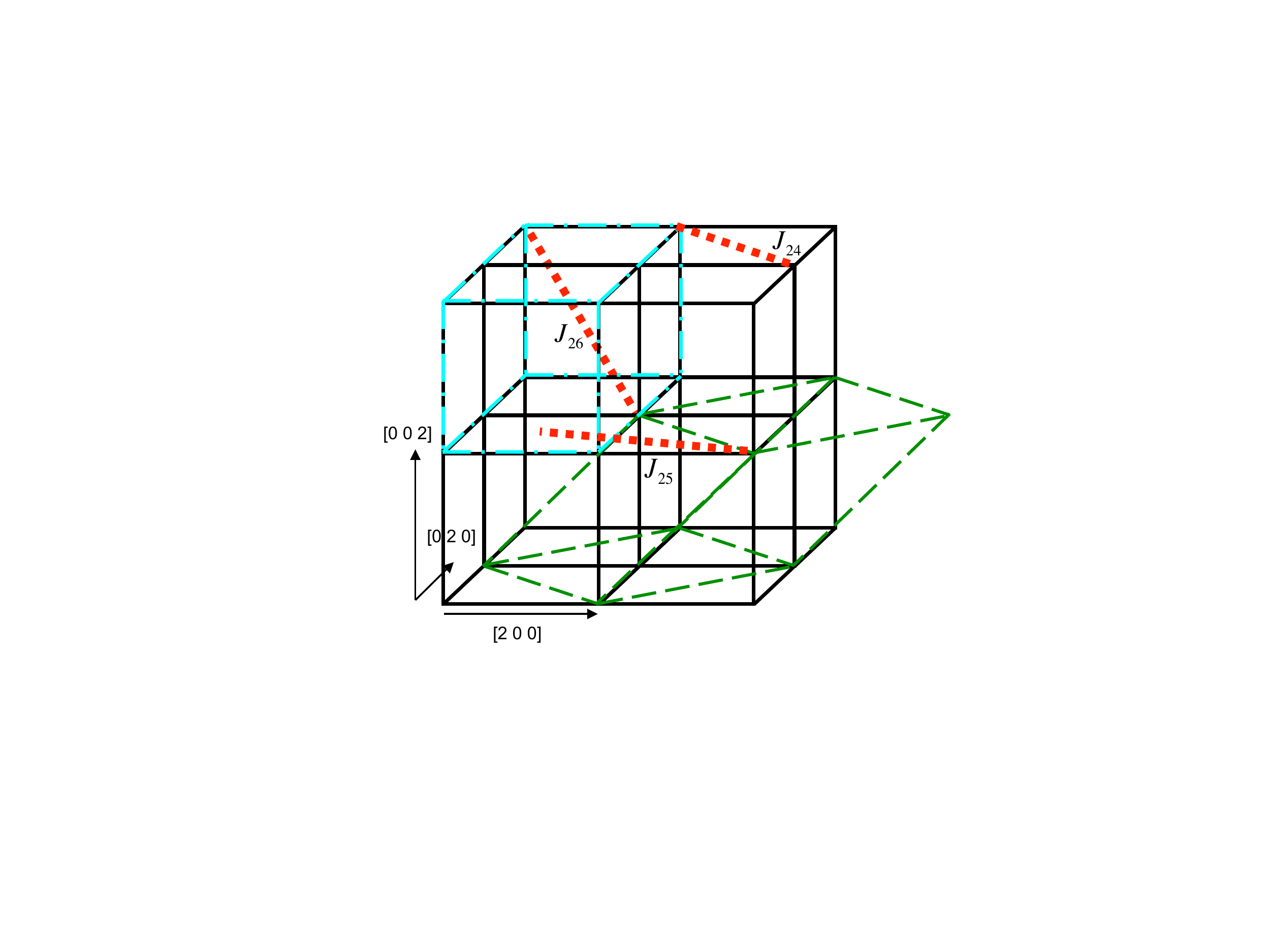} }
  \subfigure[]{
   \includegraphics[scale=0.55]{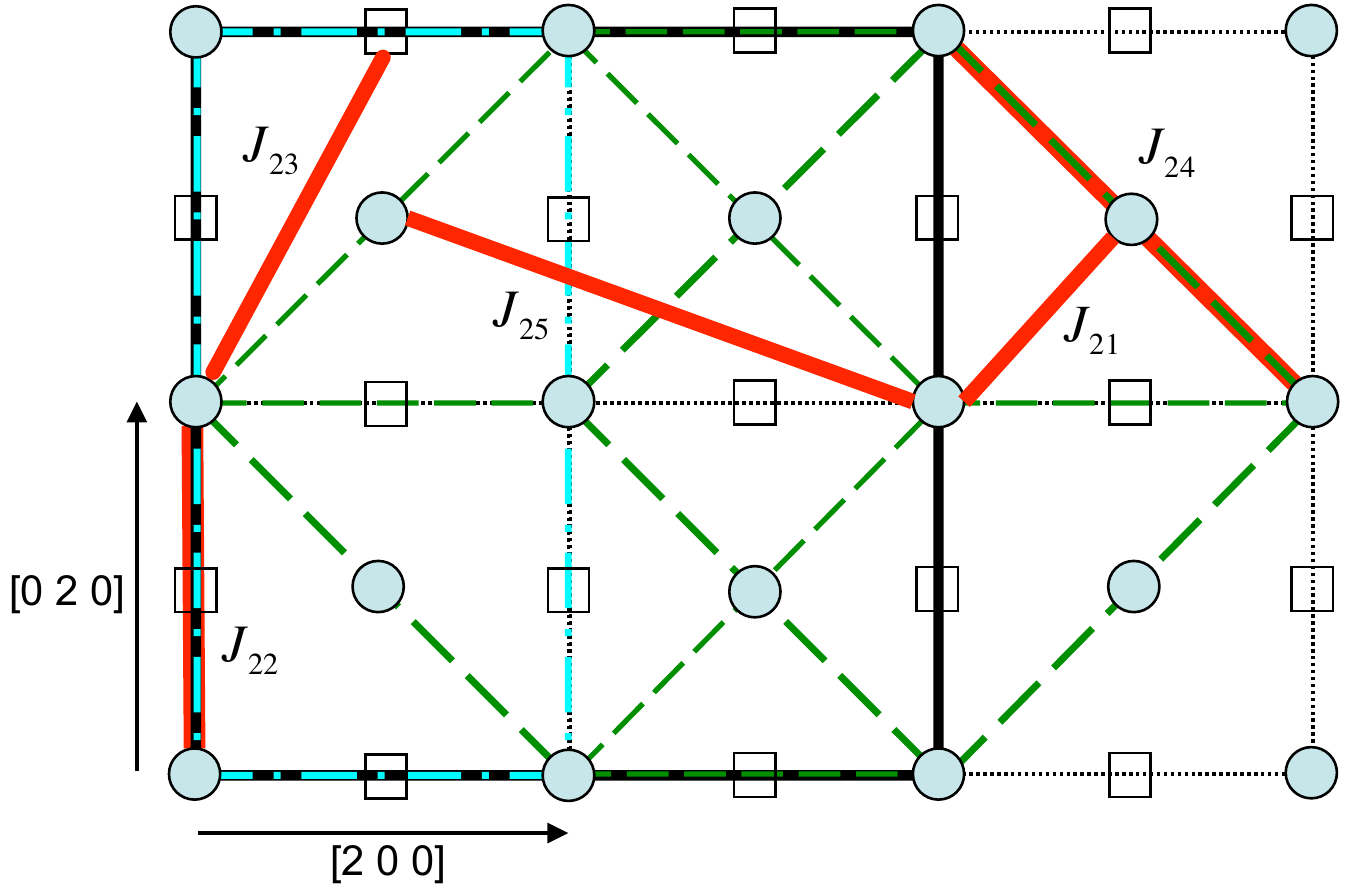} }
    \caption{\label{Fig_32_AT_space} (color online) (a) Schematic of a 32-Cubic FCC supercell forming a Hilbert space. Subspaces formed by 8-Rh (dashed) and 8-DO$_{22}$ (dot-dashed) supercells are shown. Translation vectors are given in Table \ref{Tab_subspaces}. (b) Supercells viewed along  [0 0 1], with corner (circle) and face-centered (square) sites marked. 
 For convenience, the lattice constant (corresponding to 2$^{nd}$ NN) is given 2 units. 
 For the 'complete' 32-Cubic cell, all pairs up to $J_{24}$ are unconfounded, so is $J_{26}$. However, $J_{25}$ is confounded with $J_{21}$.
For the 8-$Rh$-cell, all pairs up to $J_{22}$ are unconfounded, but $J_{23}$ is confounded with $J_{21}$.
For the 8-DO$_{22}$ subspace, pairs up to $J_{23}$ are unconfounded, but ${J}_{24}$ is confounded either with $J_0$ or ${J}_{22}$.
    }
  \end{center}
\end{figure}


\section{Augmented Subspace-Projection: FCC lattices} \label{Sec_SSP}
We now exemplify the formalism for practical application, applied to FCC Ag-Au in Section~\ref{Results}.
For FCC alloys, a cubic 32-atom supercell (Fig. \ref {Fig_32_AT_space}) is selected as a `complete' Hilbert space (denoted as 32-Cubic) with $2^{32}$ ($\approx$$4.3$ billion) configurations.
If all important ECI are unconfounded, based on the aforementioned physical hierarchy, key clusters up to a size of $n=2^{32}$ will be included.
From the moment theorem, \cite{Cyrot1968, Cyrot1970, Cyrot1971} clusters beyond a certain order should have negligible ECI for metallic alloys;  a properly trCE neglecting such terms will still predict well the  energies.
Given the CE basis set requirements stated in Sec.~\ref{subsect-hierarchy}, only a few multibodies ECI are significant generally (shown for Ag-Au in Sect.~\ref{Results}).
Thus, we can construct a CE using subspaces in the 32-Cubic space, Fig. \ref {Fig_32_AT_space}, to unconfound important multibody ECI for most alloys. 

{\par} Two 8-atom subspaces within the `complete' 32-Cubic space are considered here; the 8-Rh and the 8-DO$_{22}$ subspaces, Fig. \ref{Fig_32_AT_space}, consisting of structures generated by a symmetric rhombohedral cell and a (less symmetric) rectangular cell, respectively. 
The translation vectors of these supercells are given in Table \ref{Tab_subspaces}.
The complete space of each 8-atom subspace consists of 2$^{8}$ configurations. 
However, due to the underlying lattice symmetry and cell shape, there are only 16 and 27 unique structures for the 8-Rh and 8-DO$_{22}$ subspaces, respectively.
These two subspaces overlap, with the groundstate structures generated by the 4-Cubic space common to both.
Each of the subspace contains the usual 'suspects' for  FCC groundstate structures: \cite{AgAu2008} L1$_0$, A-rich L1$_2$, B-rich L1$_2$, pure A and B.
Low-energy configurations related to DO$_{22}$ structure are only present in 8-DO$_{22}$.
On top of multibodies beyond 1$^{st}$ NN, both the 8-Rh and 8-DO$_{22}$ spaces must necessarily unconfound $J_{01}, J_{11}, J_{21}, J_{31}, J_{41}$ because they both encompass the 4-Cubic space (see Sec. \ref{Sec_4AT_Cell}). For clarity, when we say an ECI is unconfounded, it implies that the ECI is unconfounded from lower-order (smaller $n$) and shorter-ranged ECI.

\subsection {Full Augmented Subspace-Projection: unconfounding longer-range pairs}
For an unique trCE, longer-ranged pairs are more critical than shorter-ranged multibodies.
Using the methods in Sec.~\ref{Sec_Confound}, the confounding relations for the subspaces can be worked out. 
However, as noted, the use of a larger supercell unconfounds multibody ($n>2$) clusters faster than longer-ranged pairs ($J_{2f}$ remains significant to a longer range than multibody $J_{nf}$ with $n>2$).
As it turns out, see Fig. \ref{Fig_32_AT_space}, in going from a 4-cubic cell to 8-atom cells, one only unconfounds pairs up to $J_{22}$ for the 8-Rh subspace and up to $J_{23}$ for the 8-DO$_{22}$ subspace.
On inspection of the cell geometry, we observe that even the 'complete' 32-Cubic space unconfounds only up to the $4^{th}$ NN pairs ($J_{24}$) and the $6^{th}$ NN pair ($J_{26}$) while the $5^{th}$ NN pair ($J_{25}$) remains confounded, see Fig.~\ref{Fig_32_AT_space}.
Structures from an augmented space must be added to unconfound $J_{25}$ and those beyond $6^{th}$ NN.

{\par}Our augmentation approach allows greater flexibility than the original FFD. 
Each targeted ECI is unconfounded by adding one structural energy from an augmenting space to $\vec{\text{E}}_1$, so, notably, the number of ECI in $\vec{\hat{J}}_1$ equals the number of structures in $\vec{\text{E}}_1$.
When the configuration space used is large enough to unconfound important ECI, the physical hierarchy ensures a uniquely trCE that approaches the exact one for the alloy.
Collectively, the concepts discussed and illustrated here constitute our \emph{subspace-projection} formalism.

{\par} To unconfound $J_{24}$, it suffices to combine 8-DO$_{22}$ with non-overlapping configurations from 8-Rh.
To unconfound $J_{25}$, a structure from an augmented space orthogonal to the 32-Cubic space must be added. More structures from an augmented space orthogonal to 32-Cubic are needed to unconfound longer-ranged pairs; just three more structures are required to produce an excellent CE for Ag-Au, see Sec.~\ref{Results}.

\begin{table}[]
  \begin{center}
   \caption{\label{Tab_subspaces} Translation vectors of FCC supercells representing the various spaces in Fig.~\ref{Fig_32_AT_space}, with lattice constant $a=2$. The number of sites and symmetry--unique structures generated by each supercell are listed under N$_\text{s}$ and N$_\text{c}$, respectively, with some example structures shown.
The 4-Cubic space is a subspace of 8-Rh and 8-DO$_{22}$, both of which form (overlapping) subspaces within the 32-Cubic space. N$_\text{c}$ was not evaluated exactly for the 32-Cubic, which covers a space of 2$^{32}$ non-unique configurations. The confounding relations for $n=2$ ECI up to 6$^{th}$ NN are shown too. Unless assigned the letter 'N'  (not confounded), the ECI is confounded with other ECI of higher importance (of smaller $n$ and shorter range are listed) in the particular subspace.
   }
    \begin{tabular}{c|c|c|c|c}
       Subspaces & 4-Cubic & 8-Rh & 8-DO$_{22}$ & 32-Cubic \\
        \hline
         N$_\text{s}$ & 4 & 8 & 8& 32 \\
         N$_\text{c}$ & 5 &16 & 27 &--- \\
         \hline
         Trans. vectors & [2 0 0] & [2 2 0] &[2 0 0]&[4 0 0]\\
                                    &[0 2 0] &  [2 -2 0] & [0 4 0] & [0 4 0] \\
                                    &[0 0 2] & [0 2 2] & [0 0 2]  &  [0 0 4] \\
       \hline
       Example  & Ag, Au, & 4-Cubic,  &4-Cubic,&All \\
       Structures & L1$_{0}$, L1$_{2}$ & ... & DO$_{22}$, ...  & \\
       \hline
       Confounded? &&&&\\
       $J_{21}$ & N & N & N & N \\
        $J_{22}$ & $J_{0}$ & N & N &N \\
        $J_{23}$ & $J_{21}$& $J_{21}$ & N &N \\
        $J_{24}$ &$J_0$& $J_0$ & $J_{0}, J_{22}$ & N \\
        $J_{25}$& $J_{21}$ &$J_{21}$&$J_{21}$ &$J_{21}$ \\
       $J_{26}$&$J_0$& $J_{22}$ & $J_{0}, J_{22}$ & N\\       
    \end{tabular}
  \end{center}
\end{table}

\subsection {Practical Considerations}
When complete subspaces, e.g., 8-Rh and 8-DO$_{22}$, are considered, the CE basis is locally complete and $\phi_{11}$, see \eqref{Eq_CE_Mat_break}, is directly related to the Hadamard matrix, see \eqref{Eq_CE_Mat_Sym}.
The confounding relations with longer-ranged ECI (external to the subspace) can be worked out geometrically, using concepts from FFD.
However, if only some of the structures in the subspace are used in structural inversion, $\phi_{11}$ may not be a Hadamard matrix and confounding relations between the ECI may not be determined simply from the geometry of the subspace. In such cases, we still construct  $\vec{\hat J}_1$ according to the physical hierarchy, but we check against
confounding between the ECI by ensuring that $\phi_{11}^\text{T}\phi_{11} $ is determinate.

\begin{table*}[]
\begin{center}
\caption{\label{Tab_int} ${\hat{J}}_{nf}$ (in $m$eV) and their degeneracies $D_{nf}$ for Ag-Au for different CE sets.
For  subspace projection, \{a, b, c\} represents the number of symmetry distinct  structures (see text) from 8-Rh, 8-DO$_{22}$ and augmented spaces, respectively. Left blank are ECI of clusters not used during SI, while ECI smaller than $|0.005|~$meV are listed as $\pm 0.00$.
Set \{16, 0, 0\} contains 5 other ECI (with $5\leq n \leq 8$) that are not listed as they are smaller than $|0.005|~$meV.
Set \{8, 4, 4\}* uses the same structures as \{8, 4, 4\}, but unconfounds $J_{44}$ instead of $J_{27}$ (see text).
For comparison, ECI selection via CV$_1$ score using 55-structure learning set \cite{AgAu2008} are also given.
The rms ($\varepsilon_{\text{rms}}$) and maximum ($\varepsilon_{\text{max}}$) deviation of E$^{\text{CE}}_f$ with respect to E$^{\text{DFT}}_f$ are shown with references to the figures of E$_f$ vs. \%Au. 
  }
\begin{tabular}{c|c|c|c|c|c|c|c|c|c|c|c|c}
\hline \hline
& & & \multicolumn {3} {c} {~~~~~~~~~~~~~~~~~~~~${\hat {J}}_{nf}$ ($m$eV) } \\
\hline
$n$ & $f$ & $D_{nf}$  & \{5, 0, 0\} & \{16, 0, 0\}&\{8, 0, 0\} & \{8, 4 ,0\} & \{8, 4, 1\} &\{8, 4, 2\}&\{8, 4, 4\}&\{8, 4, 6\}&CV$_1$ &\{8, 4, 4\}*\\
\hline
0 & 1 & 1  & -3010.31& -3009.22 & -3009.26 & -3008.98 &-3008.98& -3010.21 &-3010.21 &-3010.52& -3010.48&-3009.93\\
\hline
1 & 1 & 1  & -238.13&  -237.90 & -237.89 & -237.89  &-237.89 &-237.89 &-237.35 &-237.25&-237.23&-237.35\\
\hline
2 & 1 & 12  & 7.65 &~7.61  &~7.58 &~7.24&~6.59&~6.79 &~6.79&~6.90&~6.88&~7.16\\
  & 2 & 6     &           & -0.33 & -0.35  &-0.35 &-0.35&~0.06  &~0.06&~0.06&~0.10&-0.22\\
  & 3 & 24  &            &           &            &~0.17&~0.17 &~0.17&~0.36&~0.28&~0.28&~0.31\\
  & 4 & 12  &            &           &            &-0.05 &-0.05 &~0.16  &~0.16&~0.16&~0.15&~0.16\\
  & 5 & 24  &            &           &            &          &~0.33 &~0.22 &~0.22&~0.17&~0.15&~0.04\\
  & 6 & 8    &            &            &           &           &           &-0.31  &-0.31&-0.31  &-0.25 &~0.04\\
  & 7 & 48  &            &            &           &           &           &           &-0.09 &-0.05 &-0.05 &\\
  & 8 & 6    &            &            &           &           &            &           &          &~0.10&~0.12&\\
\hline
3 & 1 & 24 & -0.00 &~0.00   &-0.00 &-0.31 & -0.31  &-0.31&-0.18 &-0.18 &-0.18  & -0.18\\
  & 2 & 36  &            & -0.02   &-0.02 &-0.02 & -0.02  &-0.02&-0.02 &-0.00 &-0.06  &-0.02\\
  & 3 & 72  &            &            &            &~0.10&~0.10 &~0.10&~0.06&~0.06&~0.06&~0.06\\
  & 4 & 18  &            &            &            &           &           &           &-0.09 &-0.11&-0.05 &-0.09\\
  & 5 & 72  &            &            &            &           &           &           &           &-0.00&          &\\
\hline
4 &  1  & 8  & -0.16 &-0.07  &-0.16   & -0.07 & -0.07 &-0.07&-0.07&-0.07&-0.08&~0.77\\
  &  2  & 48 &           &~0.02  &~0.03 &~0.03 &~0.03 &~0.03&~0.03&~0.03&       &-0.11\\
  &  3  & 48 &           & -0.02  &           &-0.02  &-0.02   &-0.02& -0.02&-0.02 &       &-0.16\\
  &  4  & 12 &           & -0.03  &           &           &             &         &            &          &      &-0.28\\
 \hline \hline
\multicolumn{3}{c|}{$\varepsilon_{\text{rms}}$}&1.12  &1.08 &1.12  &0.95 &0.79 & 0.5& 0.42 & 0.42 &0.33 & 0.92\\
\multicolumn{3}{c|}{$\varepsilon_{\text{max}}$}&3.70  &4.79 &5.03 &3.90&3.28 & 1.95& 1.41 &1.22 &0.88 & 4.38\\
\multicolumn{3}{c|} {Figure}&\ref{Fig4ATFCC} & \ref{Fig8AT_only}&\ref{FigSYM8}&\ref{FigSYM8_DO224}&\ref{FigSYM8_DO224_O1} &\ref{FigSYM8_DO224_O2}&\ref{Fig_SYM8_DO224_Oth4} & ---   &\ref{Fig_CVFit_55}& \ref{Fig_844wrong}\\
\hline
\hline
\end{tabular}
\end{center}
\end{table*}

\section {Results}\label{Results}
The various CE results from subspace-projection formalism in Sec.~\ref{Sec_SSP} are showcased using structures from the 8-Rh, 8-DO$_{22}$ and augmented subspaces.
To distinguish different sets of CE, we classify the structures in each CE set by the triplet $\{a, b, c\}$;  `$a$' is the number of structures from 8-Rh, which includes all structures generated by 4-Cubic space (see Table \ref{Tab_subspaces}), `$b$' gives the number of additional structures from 8-DO$_{22}$ not found in 8-Rh, and `$c$' is the number of additional structures from the augmented space orthogonal to both 8-Rh and 8-DO$_{22}$.
The total number of structures used for direct structural inversion (SI) is $a+b+c$.

{\par}For Ag-Au alloys, we use a database of 95 DFT structural formation energies (E$^{\text{DFT}}_f$, from smallest first algorithm \cite{AgAu2008} and the subspaces above) for construction, verification and comparison of various sets of trCE. Some of the structures generated by smallest first algorithms are not within the 32-Cubic space.
The formation energy is defined as
\begin{align} \label{Eq_Ef}
  \text{E}_f(\vec{\sigma})= \text{E}(\vec{\sigma})-c(\vec{\sigma}) \text{E}(\text{Au})-(1-c(\vec{\sigma})) \text{E}(\text{Ag}) ~~,
\end{align}
with $c(\vec{\sigma})$ being the concentration of Au in the given structure defined by $\vec{\sigma}$.
The E$^{\text{DFT}}_f$ are estimated to be converged in the range of 0.2~$m$eV, also setting the lower limit for precision.

{\par} The quality of each CE set is evaluated by the root-mean-square (rms) deviation of E$^{\text{CE}}_f$ with respect to E$^{\text{DFT}}_f$ for all 95 Ag-Au structures, which includes structures not in the 32-Cubic space, i.e.,
\begin{align} \label{Eq_RMS95}
  \varepsilon_{\text{rms}} = \left[\frac{1}{95} \sum^{95}_{i=1} \left(\text{E}^{\text{DFT}}_{f}(\vec{\sigma}_i)-\text{E}^{\text{CE}}_{f}(\vec{\sigma}_i) \right)^2 \right] ^{1/2}~~.
\end{align}
Via consideration of various subspaces, we show that the unique CE set obtained up to $\sim16$ structural energies is sufficient to reproduce very well the 95 E$^{\text{DFT}}_f$ (within the convergent errors of E$^{\text{DFT}}_f$), as compared to 55 structural energies used for CV$_1$ optimal fitting.\cite{AgAu2008}

\subsection{Subspace-Projection CE}
Figure \ref{Fig_EC_SSP} shows the E$_f$ versus $c$ (at.\%Au) for various CE sets from subspace projection, with their ECI listed in Table \ref{Tab_int}. Starting with the CE set \{5, 0, 0\}, where the full 4-Cubic subspace is used, there are 5 unique structures, which incidentally are groundstate structures for Ag-Au; hence, 5 ECI (up to the NN range) are used in $\vec{\hat{J}}_1$.
From subspace projection, $\boldsymbol{\phi}_{11}$, see \eqref{Eq_SI}, is a full-ranked matrix, so the E$^{\text{DFT}}_f$ of the 5 structures are reproduced exactly by $\vec{\hat{J}}_1$.
Although the E$^{\text{DFT}}_f$ in Fig. \ref{Fig4ATFCC} for structures  on the groundstate hull are reproduced well, some other structures are not distinguished due to use of a small set of ECI.
Among ECI responsible for ordering (n$\geq2$), $J_{21}$ dominates.
The dominance by lower-order, short-range clusters is explained by the moment theorem \cite{Cyrot1968, Cyrot1970, Cyrot1971} and the electronic structure origins had been verified via direct DFT calculations.\cite{AgAu2008}

{\par} Set \{16, 0, 0\} given in Fig. \ref{Fig8AT_only} is constructed from the complete 8-Rh subspace, which unconfounds more multibody interactions; those with $n\geq5$ are negligible ($<$0.005~meV) because they are much smaller than the convergent error of E$^{\text{DFT}}_f$ data.
These negligible multibody improves the quality of the CE only marginally; the $\varepsilon_{\text{rms}}$ for E$_f$ is similar to the \{5, 0, 0\} despite having 11 more ECI.
As such, one can reduce computational cost from DFT calculations by using only a fraction of the structures in the subspace for $\vec{\text{E}}_1$, see \eqref{Eq_SI}.

{\par}We construct subset \{8, 0, 0\} with a fraction of the structures from \{16, 0, 0\} , retaining only ECI with significant magnitude (i.e., neglect  ECI with $n\geq5$) in $\vec{\hat{J}}_1$.
Although this leads to `internal' confounding between the original set of 16 ECI, 
this did not change the quality of the CE because the neglected ECI are small (negligible).
This results in only a slight increase in $\varepsilon_{\text{rms}}$ (see Fig. \ref{FigSYM8}) and a minimal change ($\sim$0.05~meV) in values of ECI (Table \ref{Tab_int}).

\subsection{Subspace-Projection + Augmentation}
{\par}  Although ECI with $2\leq n\leq4$ are significant for set \{16, 0, 0\}, this could be a result of confounding with longer-ranged but important ECI.
With the physical insight that lower-order ECI are longer range, the 8-Rh subspace does a poor job of unconfounding the longer-ranged pairs;  \{16, 0, 0\} only unconfounds up to the 2nd NN pairs and triplets.
We seek to include longer-ranged ECI (in particular, $n$=$2$) by including structures from other subspaces (augmentation).

\begin{figure*}[]
\begin{center}
\subfigure[] {\label {Fig4ATFCC}
  \includegraphics[scale=0.45]{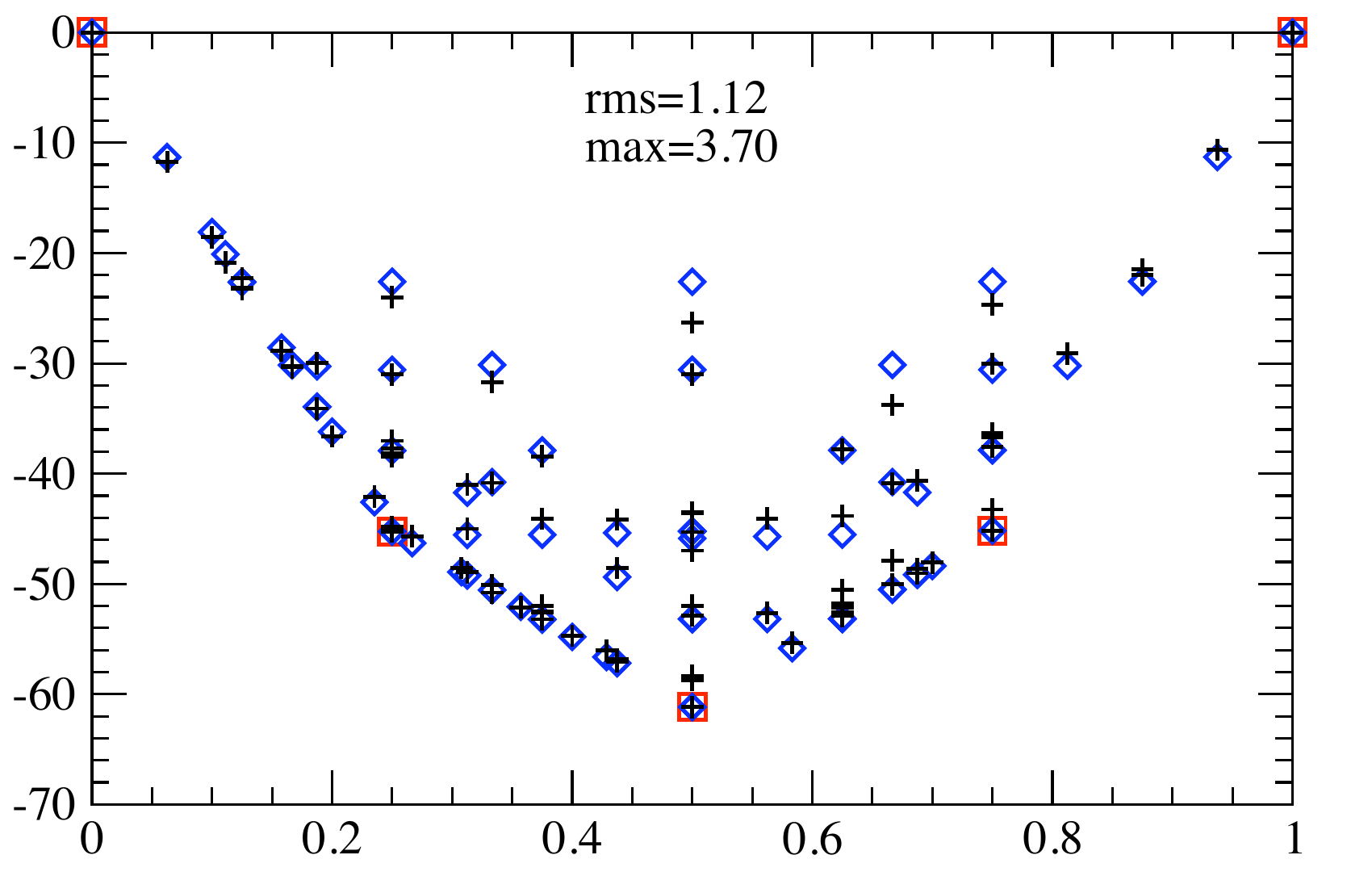}  }   
\hspace {-0.55cm}
\subfigure []{\label {Fig8AT_only}
   \includegraphics[scale=0.45]{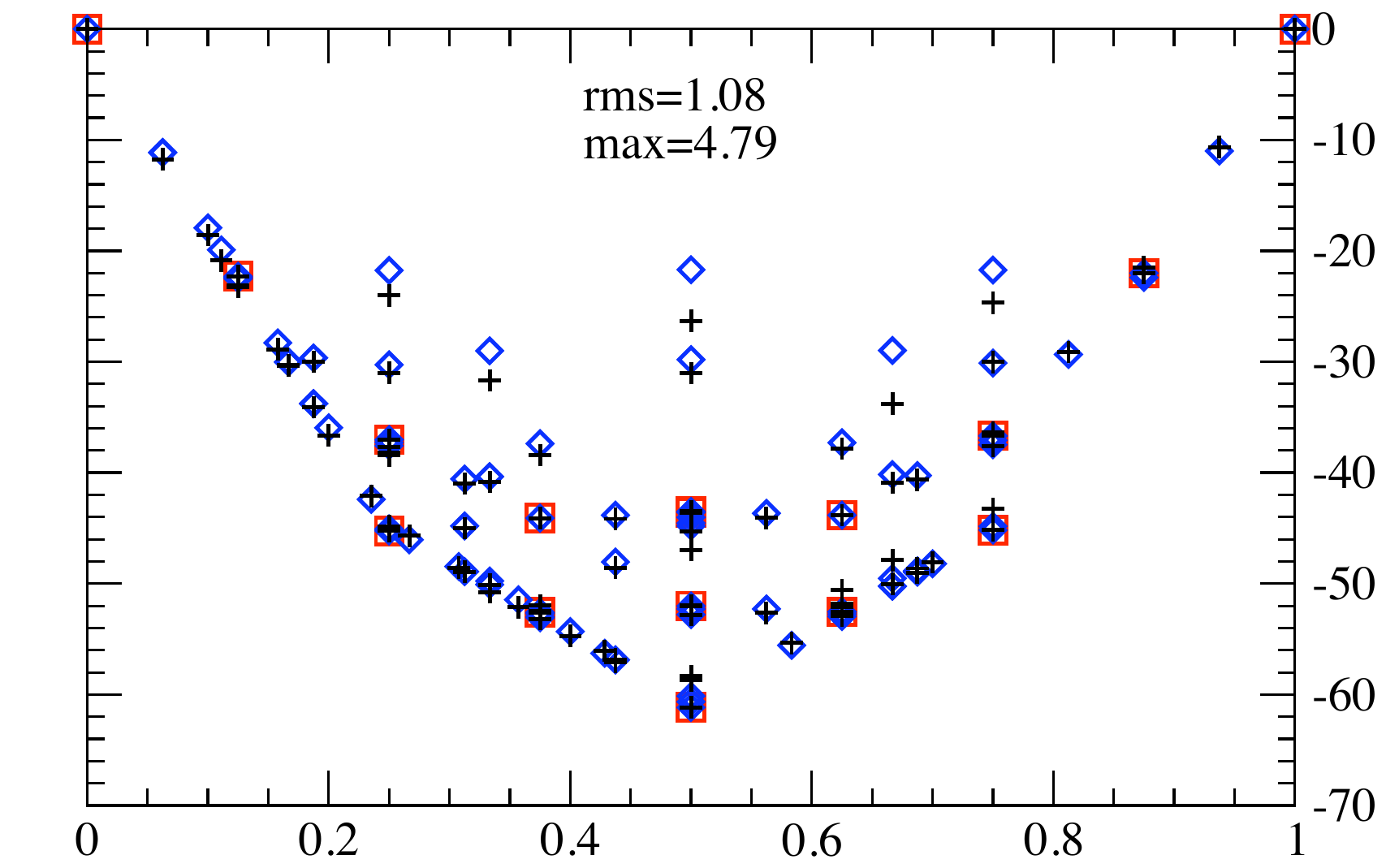}  }  
\subfigure[] {\label {FigSYM8}
  \includegraphics[scale=0.45]{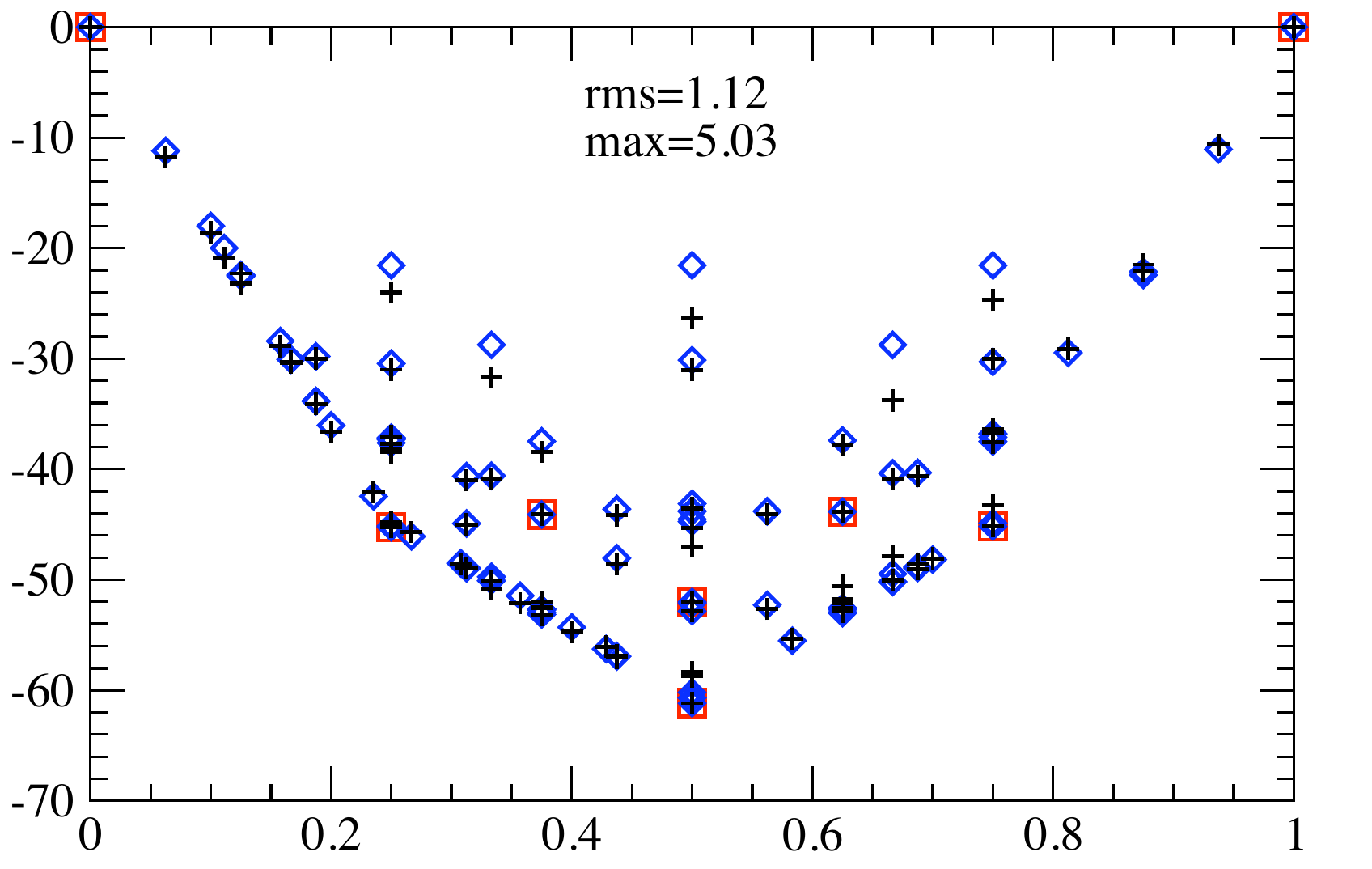}  } 
\hspace {-0.55cm}
\subfigure []{\label {FigSYM8_DO224}
   \includegraphics[scale=0.45]{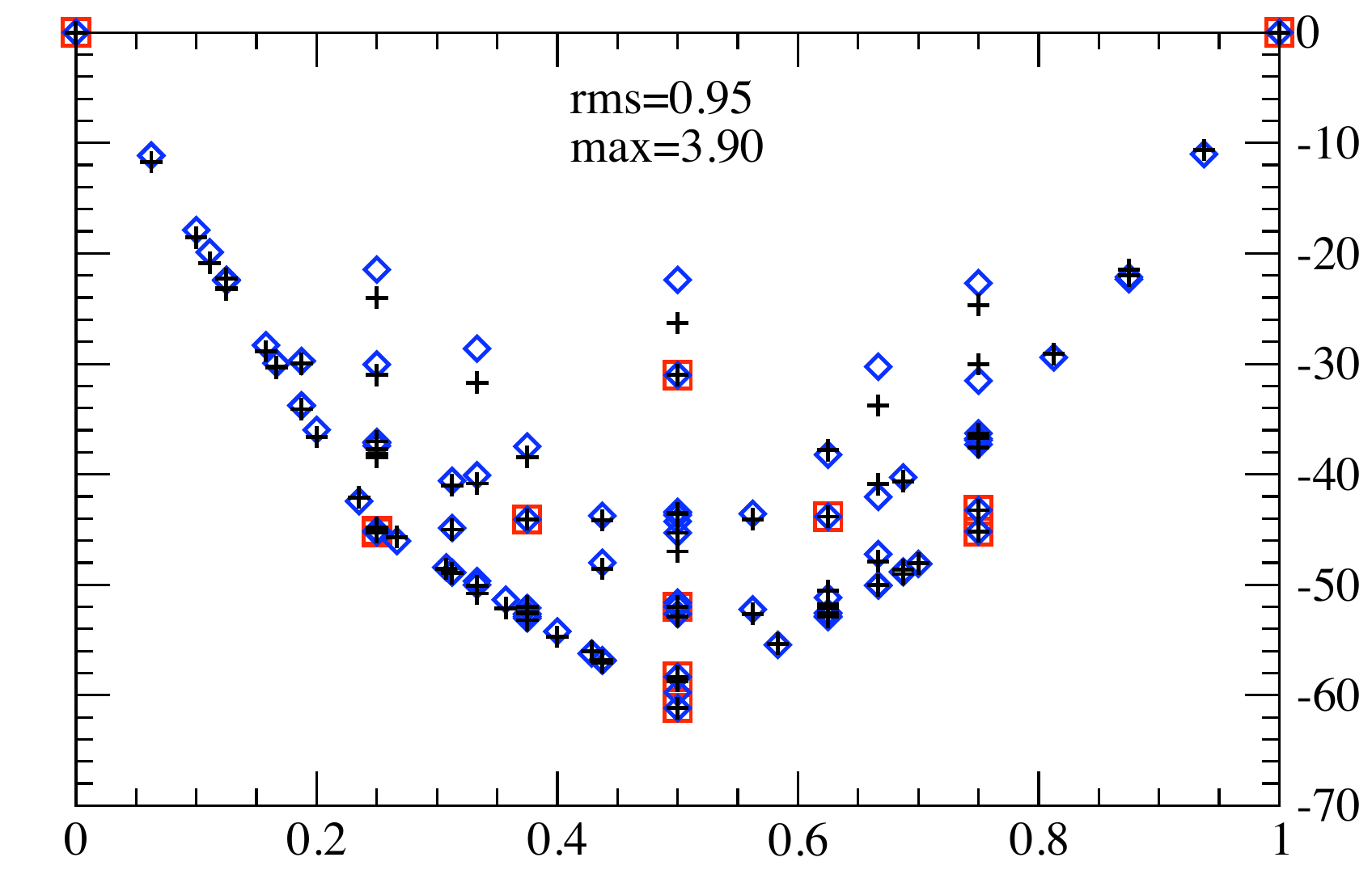}  }  
\subfigure []{\label {FigSYM8_DO224_O1}
\vspace {-10cm}
   \includegraphics[scale=0.45]{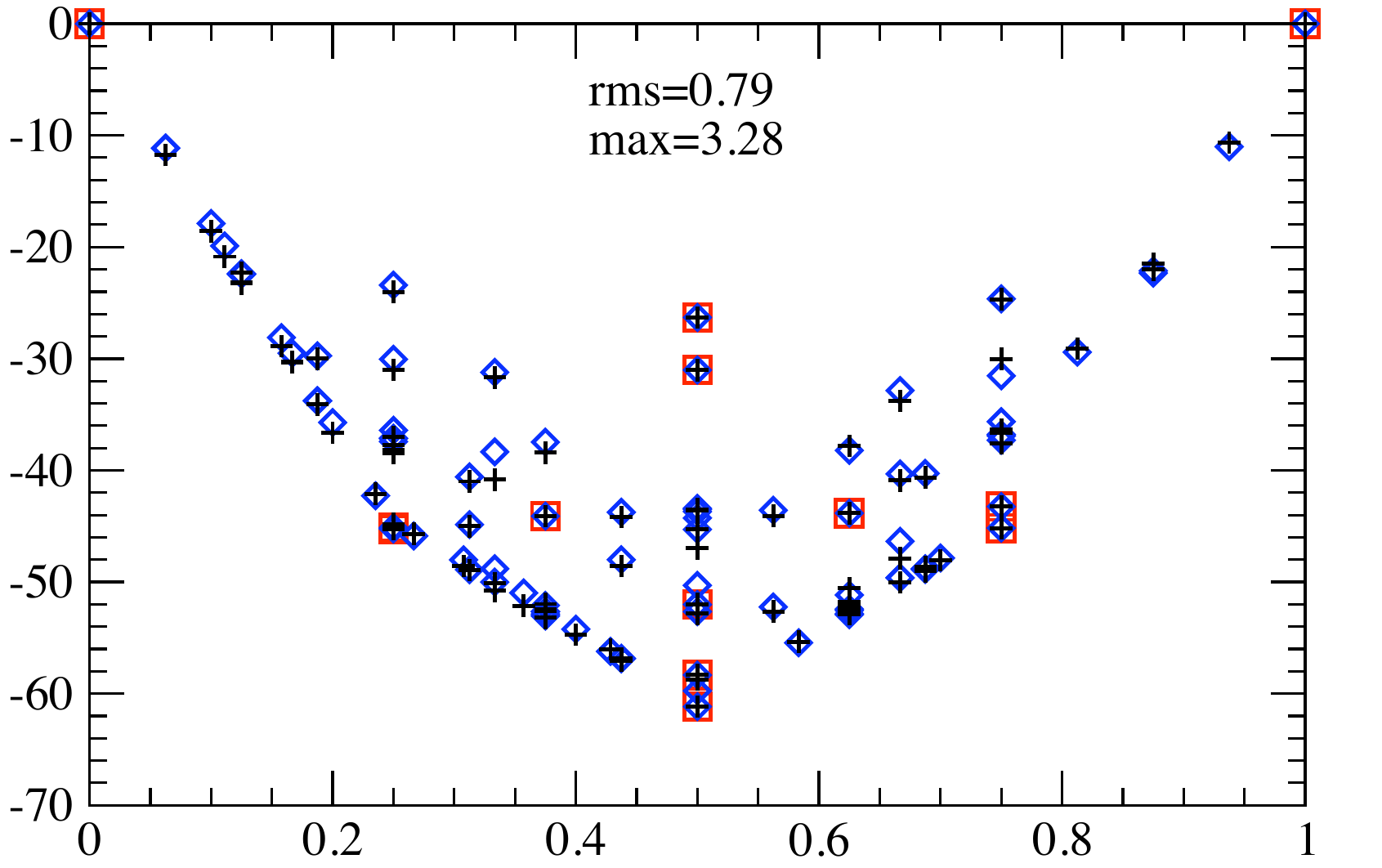}  } 
\subfigure []{\label {FigSYM8_DO224_O2}
\hspace {-0.55cm}
   \includegraphics[scale=0.45]{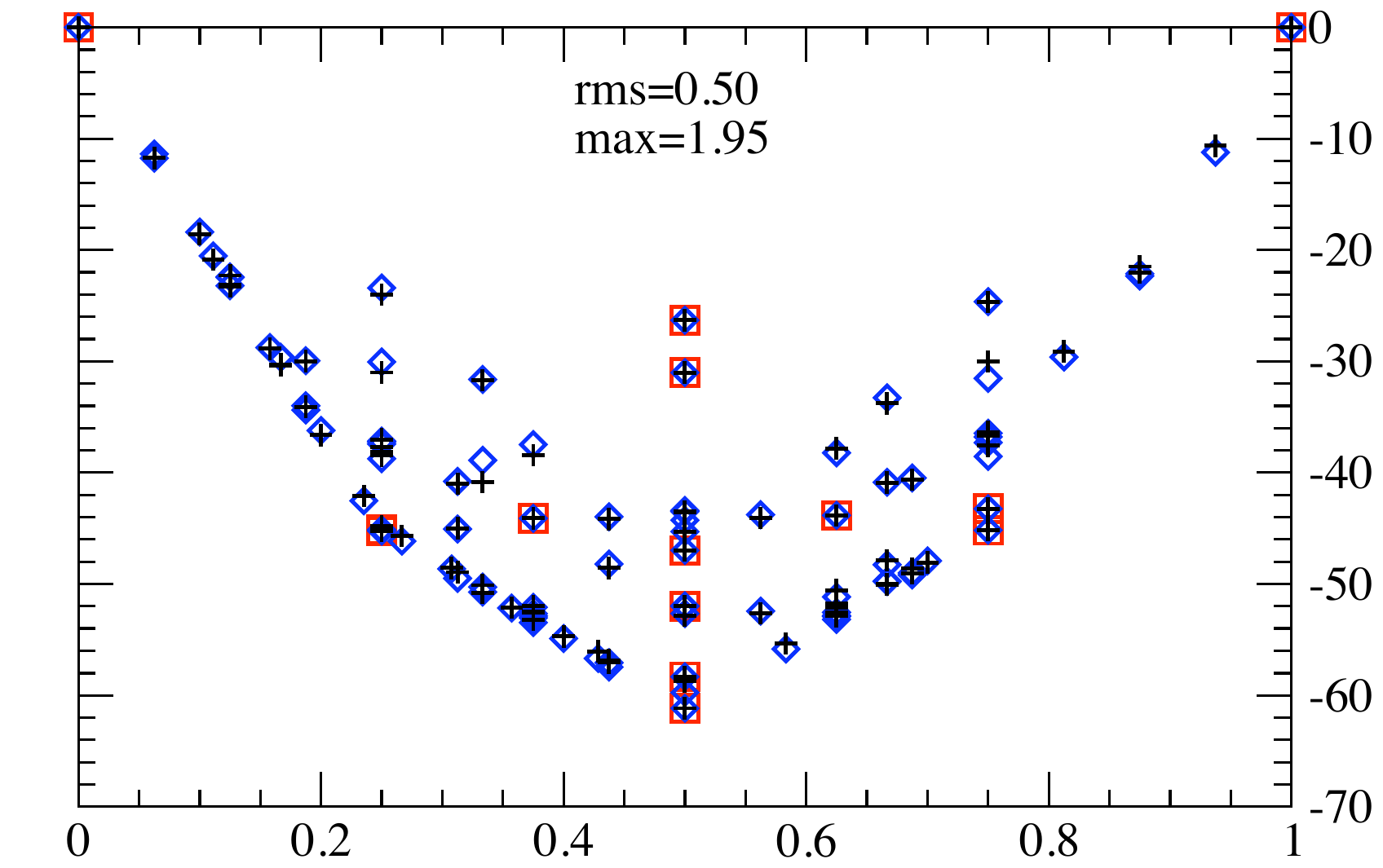}  }     
\end{center}
\vspace {-0.6cm}
\caption{\label{Fig_EC_SSP} (color online) E$^{\text{CE}}_f$ (meV) versus \%Au from CE sets (diamonds) using subspace projection from (a) \{5, 0, 0\}, (b) \{16, 0, 0\}, (c) \{8, 0, 0\}, (d) \{8, 4, 0\}, (e) \{8, 4, 1\} and (f) \{8, 4, 2\}, with  E$^{\text{DFT}}_f$ ('+') for all 95 Ag-Au structures.
Structures [(red) squares] used for structural inversion (SI) are marked.
The rms ($\varepsilon_{\text{rms}}$) and maximum ($\varepsilon_{\text{max}}$) deviation of E$^{\text{CE}}_f$ from E$^{\text{DFT}}_f$ are given for each CE set.
Only the 8-Rh subspace (including L1$_0$ and L1$_2$) are used in (a) to (c), which unconfounds up to 2$^{nd}$-NN pair at most, and they have similar $\varepsilon_{\text{rms}}$. 
(d) Adding 4 structures from 8-DO$_{22}$ cell  unconfounds the 4$^{th}$-NN pair and gives significantly better $\varepsilon_{\text{rms}}$, although high-energy structures are less well reproduced; these energies can only be improved by unconfounding the 5$^{th}$- and 6$^{th}$-NN pairs, (e) and (f), respectively, using (up to) 2 new structures from an augmented space.
}
\end{figure*}

{\par} Four structures from 8-DO$_{22}$ subspace are added to the set  \{8, 0, 0\} and the resulting \{8, 4, 0\} set unconfounds $J_{23}$, $J_{24}$, $J_{33}$ and $J_{43}$,  reducing the $\varepsilon_{\text{rms}}$ by $\sim$0.2~meV. The E$^{\text{CE}}_f$ of L1$_2$ and DO$_{22}$ are now distinguished and reproduce the E$^{\text{DFT}}_f$, see Fig. \ref{FigSYM8_DO224}.
However, the $\text{E}_f^{\text{DFT}}$ of high-energy structures are still not reproduced well, but can be improved by including pairs beyond 4$^{th}$ NN.

\begin{figure}[]
  \begin{center}
   \subfigure[]{\label {Fig_CVFit_55}
  \includegraphics[scale=0.45]{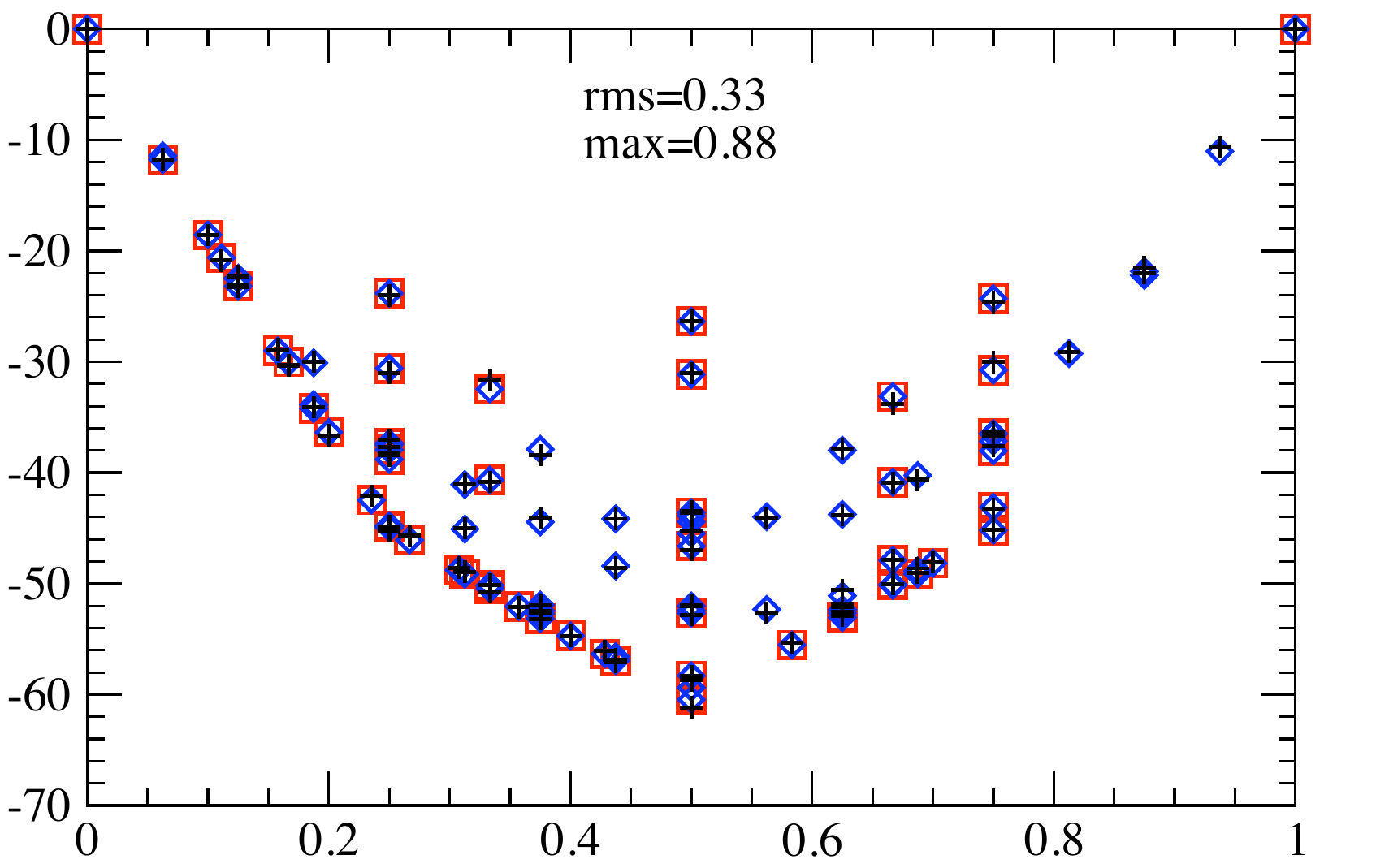} }  
  \subfigure[]{\label {Fig_SYM8_DO224_Oth4}
  \includegraphics[scale=0.45]{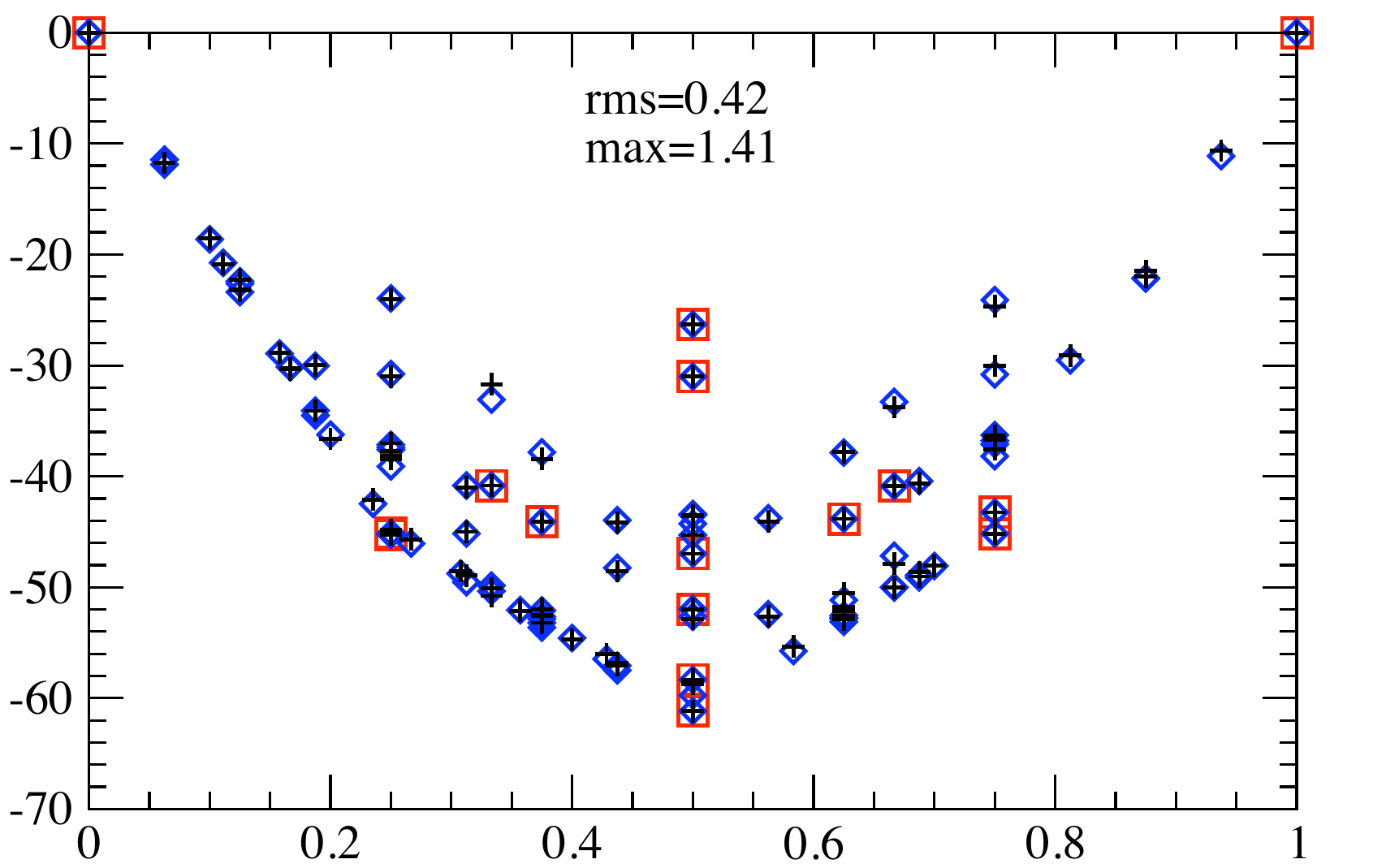} } 
  \caption{\label{Fig_fit55} (color online) E$^{\text{CE}}_f$ ($m$eV) versus \%Au (diamonds) using (a) CE selected via CV$_1$ using 55 structures (squares) and (b) CE from the \{8, 4, 4\} subspace-projection with 16 structures (squares). E$^{\text{DFT}}_f$  are denoted by '+'.
  }
\end{center}
\end{figure}

{\par} To further unconfound pairs, structures from an augmented space (see Table \ref{Tab_list_struct}) are required, because combining 8-Rh and 8-DO$_{22}$ (subspaces of 32-Cubic space) at best unconfounds  $J_{24}$.
Two structures at 0.5 Au are added in turn to give sets \{8, 4, 1\} and \{8, 4, 2\}, shown in Figs. \ref{FigSYM8_DO224_O1} and \ref{FigSYM8_DO224_O2};  unconfounding $J_{25}$ and $J_{26}$ leads to significant improvement in  $\varepsilon_{\text{rms}}$ (by $\sim$0.5~meV), which are now within the convergent errors of our E$^{\text{DFT}}_f$ data.
Unconfounding $J_{27}$ and $J_{34}$ with set \{8, 4, 4\} further reduces $\varepsilon_{\text{rms}}$ to 0.42 ~meV, see Fig. \ref{Fig_SYM8_DO224_Oth4}.

{\par} Hence, with a subset of 12 structures from the Hilbert subspace, augmented by 4 structures (having cluster functions orthogonal to that subspace) to unconfound longer-ranged pairs, an excellent quality trCE for Ag-Au with key physical ECI is found using only 16 structures.

\begin{table}[t]
  \begin{center}
   \caption{\label{Tab_list_struct} Translation vectors of 16 FCC structures (prior to atomic relaxation) used in CE set \{8, 4, 4\} with their affiliated subspaces given. The denominator in column 'Fraction Au' gives the number of atomic sites in the unit cell of each structure. Structures SM\#21, 27, 06 and 07 are from the augmented space $\perp$ to both 8-Rh and 8-DO$_{22}$ subspaces, and except for SM\#21 are also $\perp$ to the 32-Cubic space.
   }
    \begin{tabular}{c|c|c|l}
    \hline
    Tag&Fraction&Affiliated& ~~~Translation\\
           &Au          &spaces   &~~~~~vectors\\
    \hline
    Ag&  0   &All&[0 1 1], [1 0 1], [1 1 0] \\
    Au&  1   &All&[0 1 1], [1 0 1], [1 1 0] \\
    L1$_{0}$ & 1/2 & 8-Rh, 8-DO$_{22}$ &[1 1 0], [1 -1 0], [0 0 2]\\
    L1$_{2}$ & 1/4 & 8-Rh, 8-DO$_{22}$ &[2 0 0], [0 2 0], [0 0 2]\\
    L1$_{2}$ & 3/4 & 8-Rh, 8-DO$_{22}$ &[2 0 0], [0 2 0], [0 0 2]\\
    \hline
    8-Rh\#3    & 3/8 & 8-Rh                           &[2 2 0], [2 -2 0], [0 2 2]\\ 
    8-Rh\#7    & 4/8 & 8-Rh                           &[2 2 0], [2 -2 0], [0 2 2]\\ 
    8-Rh\#9    & 5/8 & 8-Rh                           &[2 2 0], [2 -2 0], [0 2 2]\\
    \hline
    DO$_{22}$ &1/4 & 8-DO$_{22}$          &[2 0 0], [0 2 0], [1 1 2]\\
    DO$_{22}$ &3/4 & 8-DO$_{22}$          &[2 0 0], [0 2 0], [1 1 2]\\
    SM\#13 &2/4 & 8-DO$_{22}$          &[2 0 0], [0 2 0], [1 1 2]\\
    SM\#24 &2/4 & 8-DO$_{22}$          &[4 0 0], [0 1 -1], [0 1 1]\\
    \hline
    SM\#21 &2/4 & Aug., in 32-Cubic     &[1 -1 0], [2 2 0],  [0 0 2 ]\\
    SM\#27 &2/4 & Aug., $\perp$ 32-Cubic  & [3 3 2], [0 1 -1], [-1 0 1]\\
    SM\#06    &1/3 & Aug., $\perp$ 32-Cubic  & [1 1 0], [1 -1 0], [1 0 3]\\
    SM\#07    &2/3 & Aug., $\perp$ 32-Cubic & [1 1 0], [1 -1 0], [1 0 3]\\
    \hline
    \end{tabular}
  \end{center}
\end{table}

\begin{figure}[t]
  \begin{center}
    \includegraphics[scale=0.45]{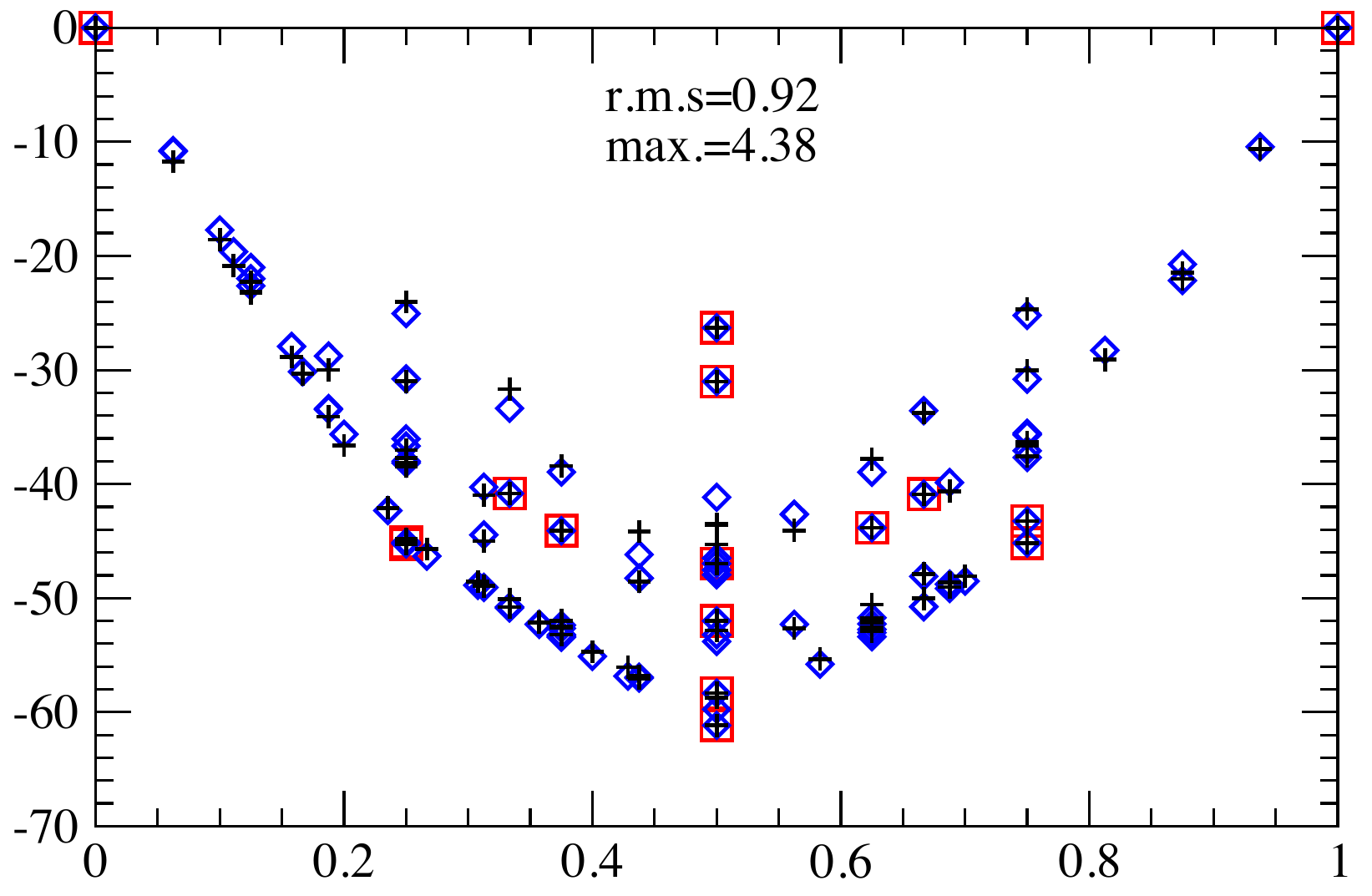} 
     \caption{\label{Fig_844wrong} (color online) E$^{\text{CE}}_f$ ($m$eV) versus \%Au (diamonds) using CE set \{8, 4, 4\}* where $J_{44}$ is added without observing the physical cluster hierarchy. E$^{\text{DFT}}_f$  are denoted by '+'.}
  \end{center}
\end{figure}  

\section {Discussion}\label{Discussion}
Below we discuss the relationship to and comparison with standard statistical fitting methods, and recently suggested regularization using Bayesian concepts. 

\subsection{Subspace-projection versus CV$_1$ Fitting}
We now compare  the \{8, 4, 4\} subspace-projection trCE  with the trCE obtained by minimizing CV$_1$, \cite{AgAu2008} which uses at least 55 structures (not necessarily from the 32-Cubic space) as the learning set. 
Unlike subspace-projection which used 16 structures for direct SI,  CV$_1$ selects a set via a statistical fit and is allowed to have fewer ECI than the number of DFT energies used for SI.
We emphasize that our CV$_1$ selection also uses the same hierarchy of clusters \cite{PRL92p55702} to ensure a locally complete CE, unlike others. \cite{Hart2005}
The small improvement of $\varepsilon_{\text{rms}}$ by 0.1 $m$eV for the optimal CV$_1$ set is a result of using more than $3$ times the structures in the learning set; that is,  the \emph{least-squares} error is minimized in \eqref{Eq_SI} over 55 structures, which is a large fraction of the 95 structures used for validation  by $\varepsilon_{\text{rms}}$ in \eqref{Eq_RMS95}.
So, it is not surprising that there is a slight improvement using CV$_1$, because the ECI values are altered to improve the fit.

{\par} To facilitate comparison of the ECIs, we further construct set  \{8, 4, 6\}, unconfounding $J_{28}$ and $J_{35}$.
The improvement in $\varepsilon_{\text{rms}}$ is insignificant versus \{8, 4, 4\}.
As observed in Table \ref{Tab_int}, the ECI of \{8, 4, 4\}, \{8, 4, 6\} and CV$_1$ (55-structure) fit are very close to one another, signifying a convergence in ECI, within errors of E$^{\text{DFT}}_f$.

{\par} We see that the selection of ECI based on a physical hierarchy is of primary importance, because once the physically important ECI are unconfounded, the exact CE of the alloy system is approached. 
At this point, the ECI and the accuracy of the trCE are similar regardless of the number of structural energies used in the learning set.
For example, in \{8, 4, 4\} subspace, without the physical hierarchy, one could have unconfounded $J_{44}$ instead of the physically more important $J_{27}$ (\{8, 4, 4\}* versus \{8, 4, 4\}, respectively, in Table \ref{Tab_int}), producing a CE with worse predictive capability, see Fig. \ref{Fig_844wrong}.
\emph{The bottom line:} ECI have  physical meanings and they should not be treated merely as coefficients obtained from statistical fitting.

\subsection{Relation to Bayesian Approaches}
The physical hierarchy of clusters utilized in the present paper for unconfounding (also used in our previous CV$_1$ CE\cite{PRL92p55702}) would modify the usually assumed ``uniform'' (i.e., otherwise uninformative) \emph{prior} distribution for the ECI, $\vec{J}$, within the Bayesian framework. 
The \emph{posterior} probability of $\vec{J}$ given $\vec{E}_1$ is\cite{PhysRevB.80.024103,ECockayne_2010}
\begin{align}
  P(\vec{J} | \vec{E}_1) \propto P(\vec{E}_1 | \vec{J})P^0(\vec{J})~~,
\end{align}
where $\vec{J} \equiv [\vec{J}_{1},\vec{J}_{2}]^{T}$.
$\vec{J}$ contains the truncated (non-zero) $\vec{J}_{1}$ from SI and excluded (possibly zero) $\vec{J}_{2}$.
Here, $P^0(\vec{J})$ is the \emph{prior} distribution, which is non-zero only for trCE whose ECIs are locally complete and follow the physical hierarchy in our subspace-projection CE.
In contrast, a uniformly distributed  $P^0(\vec{J})$  assumes all trCE are possible, regardless of being physical or not.

{\par} Recently, by an assumption that the ECI of a given cluster results from a large number of \emph{a priori} random contributions, a Gaussian prior distribution was proposed \cite{ECockayne_2010}; additionally, a decaying weight was assumed to cutoff smoothly contributions from ECI that otherwise are assumed zero. 
To be clear, our confounding relations, see, e.g., \eqref{Eq_BC_Cell_3}, reflect mathematically the specific  ECI in $\vec{J}_{2}$ (albeit with \emph{a priori} unknown values) that directly affect those in $\vec{J}_{1}$.
Our \emph{a priori} choice can be to set all $\vec{J}_{2}$ to zero and validate using structural energies not in the learning set. 
(\emph{A posteriori} we can augment the subspace to systematically unconfound.)
Or we can assume that $ P^0(\vec{J})$ decays according to some specifically chosen distribution,\cite{PhysRevB.80.024103,ECockayne_2010}
which certainly may be included in the present formalism.

\section{Conclusion}
To construct a unique truncated CE, we presented an {Augmented Subspace-Projection} formalism using the mathematics of Hilbert spaces and concepts from {Fractional Factorial Design (FFD)} that directly select the critical, \textit{a priori} unknown ECI in the included set of $\vec J_1$ (with excluded ones in $\vec J_2$).
As exemplified for binary alloys with an N-site lattice and Hilbert space of 2$^\text{N}$ configurations, structural energies can be reproduced by an estimator $\vec{\hat{J}}_1$, containing a minimal set of physical ECI that approaches the exact ECI.
When N is large, DFT calculations are feasible only for a vanishingly small fraction of the $2^\text{N}$ structures, resulting in linear dependencies between basis functions such that $\vec{\hat{J}}_1$ is confounded with specific ECI in $\vec{\hat{J}}_2$.
However, from FFD concepts, this confounding between ECI can be determined, so only a few ($\sim$16) structures are needed to construct, without fitting, a reliable CE with quantifiable errors, see \eqref{Eq_Hada2_confound}--\eqref{Eq_Hada2_part3}.
Of course, no statistical fitting does not imply no statistical validation.

{\par} For practical applications using structures with periodic boundary conditions, we showed that the confounding relations between ECI can be identified geometrically when subspaces (chosen supercells that lie within the defined  Hilbert space) are considered. 
A physical hierarchy of ECI provides a condition to obtain a physical set of truncated ECI.
Although the CE from the subspaces are complete, longer-ranged pairs can remain confounded with truncated ECI, which can be unconfounded by augmenting the subspace.

{\par} Using FCC Ag-Au as a case study, we defined an initial subspace by an 8-atom rhombohedral cell (8 structures), which is then augmented to construct a unique truncated CE. 
This \emph{augmented subspace-projection} formalism using 16 structures, without fitting, produces a CE with similar predictive capability as that obtained from a CV$_1$ statistical fit using $>$55 structures.
The concepts discussed above can be generalized to multicomponent alloys.

\vspace{0.5cm}
{\bf Acknowledgements:} Partial support for TLT at Illinois was from the National Science Foundation  (DMR-07-05089) for Thermodynamic Tool Kit (TTK) software.
Additional support was by the U.S. Department of Energy, Office of Basic Energy Sciences, Division of Materials Science and Engineering (DE-FG02-03ER4606) and Ames Laboratory. Ames Laboratory is operated for the US DOE by Iowa State University under contract (DE-AC02-07CH11358).
TLT acknowledges support from Institute of High Performance Computing, Singapore.

\appendix
\section {Derivation of Error Terms} \label{Appendix}
We derive the decomposition of the MSE into variance and bias terms shown in Section \ref{Sec_SI}.
\begin{align}
\text{MSE} &=  \left< \left(  \hat{\text{E}}(\vec{\sigma})-\text{E}(\vec{\sigma}) \right )^2  \right> \nonumber \\
&=\left< \left( \hat{\text{E}}(\vec{\sigma})- \left<\hat{\text{E}}(\vec{\sigma})\right>+ \left<\hat{\text{E}}(\vec{\sigma})\right>-\text{E}(\vec{\sigma}) \right)^2  \right> \nonumber \\
&=   \left<\left( \hat{\text{E}}(\vec{\sigma})- \left<\hat{\text{E}}(\vec{\sigma})\right>\right)^2\right>+ \left<\left(\left<\hat{\text{E}}(\vec{\sigma})\right>-\text{E}(\vec{\sigma}) \right)^2  \right> \nonumber \\
&~~+ 2\left< \left( \hat{\text{E}}(\vec{\sigma})- \left<\hat{\text{E}}(\vec{\sigma})\right>\right)\left(\left<\hat{\text{E}}(\vec{\sigma})\right>-\text{E}(\vec{\sigma}) \right) \right> \nonumber \\
&= \text{Var}+\text{Bias}+2~\text{Cross}~~.
\end{align}
where $<...>$ denotes expectation values averaged over all possible observations having the same atomic configuration, $\vec{\sigma}$. 

\begin{align}
\text{Cross}&= \left<\hat{\text{E}}(\vec{\sigma})\left<\hat{\text{E}}(\vec{\sigma})\right>\right> -\left< \hat{\text{E}}(\vec{\sigma})\text{E}(\vec{\sigma})\right> \nonumber \\
&~~-\left<\left<\hat{\text{E}}(\vec{\sigma})\right>\left<\hat{\text{E}}(\vec{\sigma})\right>\right>+ \left<\left<\hat{\text{E}}(\vec{\sigma})\right>\text{E}(\vec{\sigma})\right> \nonumber \\
&=\left<\hat{\text{E}}(\vec{\sigma})\right>\left<\hat{\text{E}}(\vec{\sigma})\right> - \left< \hat{\text{E}}(\vec{\sigma})\right>\text{E}(\vec{\sigma})\nonumber \\
&~~-\left<\hat{\text{E}}(\vec{\sigma})\right>\left<\hat{\text{E}}(\vec{\sigma})\right>+\left< \hat{\text{E}}(\vec{\sigma})\right>\text{E}(\vec{\sigma}) =0~~.
\end{align}
We have used the fact that $<<\hat{\text{E}}(\vec{\sigma})>>=<\hat{\text{E}}(\vec{\sigma})>$ and $<\text{E}(\vec{\sigma})>=\text{E}(\vec{\sigma})$. For the variance term,
\begin{align}
  \text{Var}&= \left<\left( \hat{\text{E}}(\vec{\sigma})- \left<\hat{\text{E}}(\vec{\sigma})\right>\right)^2\right>\nonumber \\
  &= \left<  \left[{\phi}_{R1}^{\vec{\sigma}}\left(\boldsymbol{\phi}^\text{T}_{11}\boldsymbol{\phi}_{11}\right)^{-1} \boldsymbol{\phi}^\text{T}_{11}\left(\vec{\mathcal{E}}_1-\left<\vec{\mathcal{E}}_1 \right>\right)\right]^2 \right> \nonumber \\
 &= \left<  \left[{\phi}_{R1}^{\vec{\sigma}}\left(\boldsymbol{\phi}^\text{T}_{11}\boldsymbol{\phi}_{11}\right)^{-1} \boldsymbol{\phi}^\text{T}_{11} \vec{\epsilon}_1\right]^2 \right> s^2 \nonumber \\
 &={\phi}_{R1}^{\vec{\sigma}}\left(\boldsymbol{\phi}^\text{T}_{11}\boldsymbol{\phi}_{11}\right)^{-1} {\phi}_{R1}^{\vec{\sigma}~\text{T}} s^2~~,
\end{align}
where $s^2$ is the variance of the randomly distributed error $\vec{\epsilon}$ (see \eqref{Eq_Noise}) .
For the bias term,
\begin{align}
\text{Bias}&=\left<\left(\left<\hat{\text{E}}(\vec{\sigma})\right>-\text{E}(\vec{\sigma}) \right)^2  \right> \nonumber \\
&=\left(\left<\hat{\text{E}}(\vec{\sigma})\right>-\text{E}(\vec{\sigma}) \right)^2 \nonumber \\
&=\left(  {\phi}_{R1}^{\vec{\sigma}}  \vec{\hat{J}}_1 - {\phi}_{R1}^{\vec{\sigma}}  \vec{{J}}_1 -  {\phi}_{R2}^{\vec{\sigma}} \vec{{J}}_2    \right)^2~~.
\end{align}

\newpage

\bibliography{CE_thesis}
\end{document}